\documentclass[aps,pra,twocolumn,superscriptaddress]{revtex4-2}
\usepackage{amsmath,amssymb,amsfonts,bm,physics,latexsym,comment} 
\usepackage[pdftex]{graphicx,color}
\usepackage{hyperref}
\hypersetup{
   colorlinks=true,        % false: boxed links; true: colored links
   linkcolor=blue,          % color of internal links (change box color with linkbordercolor)
   citecolor=blue,        % color of links to bibliography
  filecolor=magenta,      % color of file links
   urlcolor=blue           % color of external links
}
\usepackage{tikz}

\newcommand{\tred}[1]{\textcolor{red}{#1}}

\newcommand{\Jperp}{J_\perp}
\newcommand{\Jdiag}{J_\times}

\newcommand{\gt}{\tilde{g}}

\newcommand{\Sv}{\bm{S}}
\newcommand{\Jv}{\bm{J}}
\newcommand{\Tv}{\bm{T}}

\newcommand{\Ucal}{{\cal U}}
\newcommand{\Ucals}{\Ucal_{\rm s}}

\newcommand{\ab}{\bar{a}}
\newcommand{\db}{\bar{d}}

\newcommand{\Kch}{K_{\rm ch}}

\newcommand{\ua}{\uparrow}
\newcommand{\da}{\downarrow}

\begin{document}
%%%%%%%%%%%%%%%%%%%%%%%%%%%%%%%%%%%%%%%%%%%%%%%%%
% Paper Information
%%%%%%%%%%%%%%%%%%%%%%%%%%%%%%%%%%%%%%%%%%%%%%%%%
\title{
%Spin-chirality interplay in a spin-$\frac12$ Heisenberg ladder with a four-spin interaction
Phase Diagram of a Spin-$\frac12$ Heisenberg Ladder with a Chirality-Chirality Interaction
}
\author{Mat\'eo Fontaine}
%\homepage[ORCID: ]{https://orcid.org/0000-0002-0883-4355}
\affiliation{Department of Physics, Keio University, 3-14-1 Hiyoshi, Kohoku-ku, Yokohama, Kanagawa 223-8522, Japan}
\author{Shunsuke Furukawa}
%\homepage[ORCID: ]{https://orcid.org/0000-0003-3376-7198}
\affiliation{Department of Physics, Keio University, 3-14-1 Hiyoshi, Kohoku-ku, Yokohama, Kanagawa 223-8522, Japan}
\affiliation{Keio University Sustainable Quantum Artificial Intelligence Center (KSQAIC), Keio University, 2-15-45 Mita, Minato-ku, Tokyo 108-8345, Japan}

\date{\today}

%%%%%%%%%%%%%%%%%%%%%%%%%%%%%%%%%%%%%%%%%%%%%%%%
% Abstract
%%%%%%%%%%%%%%%%%%%%%%%%%%%%%%%%%%%%%%%%%%%%%%%%
\begin{abstract}
A spin-$\frac12$ Heisenberg model with a chirality-chirality interaction (CCI) on a two-leg ladder 
provides a minimal setup to explore an interplay between spin and chirality degrees of freedom. 
This model is potentially relevant for understanding the effects of a four-spin ring exchange and a phonon-mediated interaction on a ladder. 
By applying the Abelian bosonization and two types of spin-chirality duality, we derive a comprehensive ground-state phase diagram 
that includes previously unexplored cases of a ferromagnetic rung exchange $\Jperp<0$ or a negative CCI $K_{\rm ch}<0$. 
The phase diagram consists of the rung singlet (RS) phase, the Haldane phase, and three non-magnetic ordered phases. 
Moreover, the RS and Haldane phases consist of several regions with different dominant correlations in terms of spin, dimer, or chirality. 
We perform numerical simulations to confirm the predicted phase structure and critical properties. 
\end{abstract}
\maketitle

%%%%%%%%%%%%%%%%%%%%%%%%%%%%%%%%%%%%%%%%%%%%%%%%
% Main text
%%%%%%%%%%%%%%%%%%%%%%%%%%%%%%%%%%%%%%%%%%%%%%%%

%%%%%%%%%%%%%%%%%%%%%%%%%%%%%%%%%%%%%%%%%%%%%%%%
\section{Introduction}\label{sec:intro}
%%%%%%%%%%%%%%%%%%%%%%%%%%%%%%%%%%%%%%%%%%%%%%%%

% [ Ring exchage ]--------------------------------------
Multiple-spin exchange interactions among more than two spins have been studied intensively due to their experimental relevances 
as well as their remarkable ability to induce exotic quantum phases. 
A four-spin ring exchange appears in the fourth-order perturbation theory of the Hubbard model \cite{Takahashi77,MacDonald90}, 
and its effects on excitation spectra have been investigated 
for cuprate layers \cite{MacDonald90,RogerDelrieu89,Chubukov92,Coldea01} and ladders \cite{Brehmer99, Matsuda00, Nunner02, Schmidt05}. 
In these cuprates, the four-spin exchange coupling is estimated to be about 20\%
of the dominant Heisenberg two-spin exchange couplings, 
and its effects are mostly perturbative. 
Meanwhile, multiple-spin exchanges can play a more significant role 
in solid $^3$He \cite{Roger83} and Wigner crystals \cite{Chakravarty99, Klironomos07}. 
In particular, for the second layer of $^3$He on graphite, it has been argued that 
multiple-spin exchanges involving up to six spins have comparable magnitudes,  
and are responsible for spin-liquid behavior observed experimentally \cite{Fukuyama08, Ishida97,Masutomi04,Misguich98,Misguich99}.  
Theoretically, a four-spin ring exchange has been found to induce a variety of exotic phases, 
such as staggered dimer (SD) and scalar chiral (SC) orders on the ladder 
\cite{Laeuchli03, Hikihara03, Hijii02, Hijii03, Mueller02,  Capponi13, Gritsev04, Totsuka12, Metavitsiadis17, Furukawa06, Furukawa07}, 
a vector chiral (VC) order on the square lattice \cite{Laeuchli05,WangChao22},
and possible gapped \cite{Misguich98,Misguich99,Cookmeyer21} and gapless \cite{LiMing00,Motrunich05} quantum spin liquids on the triangular lattice.  

%cuprate layer: La2CuO4
%cuprate ladder: La$_6$Ca$_8$Cu$_{24}$O$_{41}$ 
%Other refs: Bernu01, Katano00

\begin{figure}
\includegraphics[width=0.4\textwidth]{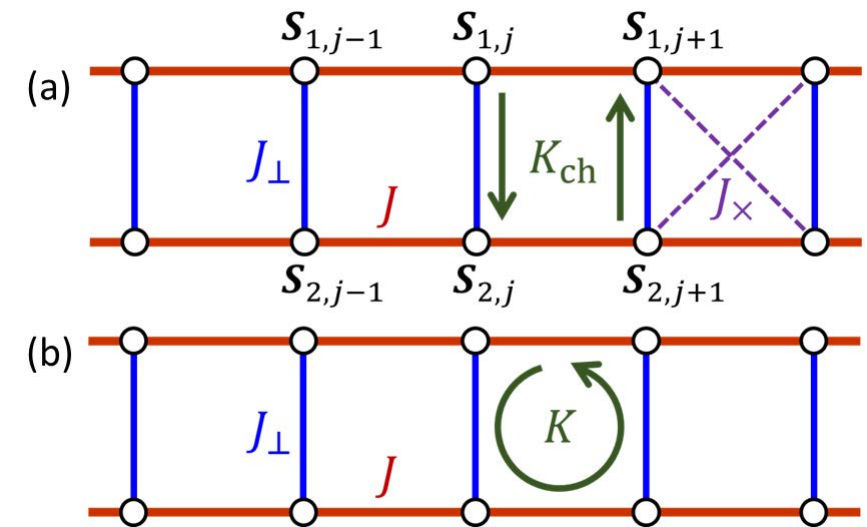}
\caption{\label{fig:laddermodel_chpm}
(Color online) 
(a) Spin-$\frac12$ Heisenberg ladder with a CCI $K_{\rm ch}$, 
described by the Hamiltonian \eqref{eq:H_XXX_CCI}. 
The spin-$\frac12$ operator at the $j$th site on the $n$th leg is denoted by $\Sv_{n,j}$. 
We set $J=1$ as the unit of energy when presenting numerical results. 
The diagonal exchange interaction $J_\times$ is also considered in the extended model \eqref{eq:H_XXX_chex}. 
(b) Similar ladder model with a four-spin ring exchange $K$ studied in 
Refs.\ \cite{Brehmer99, Matsuda00, Nunner02, Schmidt05, Laeuchli03, Hikihara03, Mueller02, Gritsev04, Hijii02, Hijii03, Hikihara08, Totsuka12, Capponi13, Metavitsiadis17, Furukawa06, Furukawa07,Hakovyan08}. 
}
\end{figure}

\begin{figure}
\includegraphics[width=0.5\textwidth]{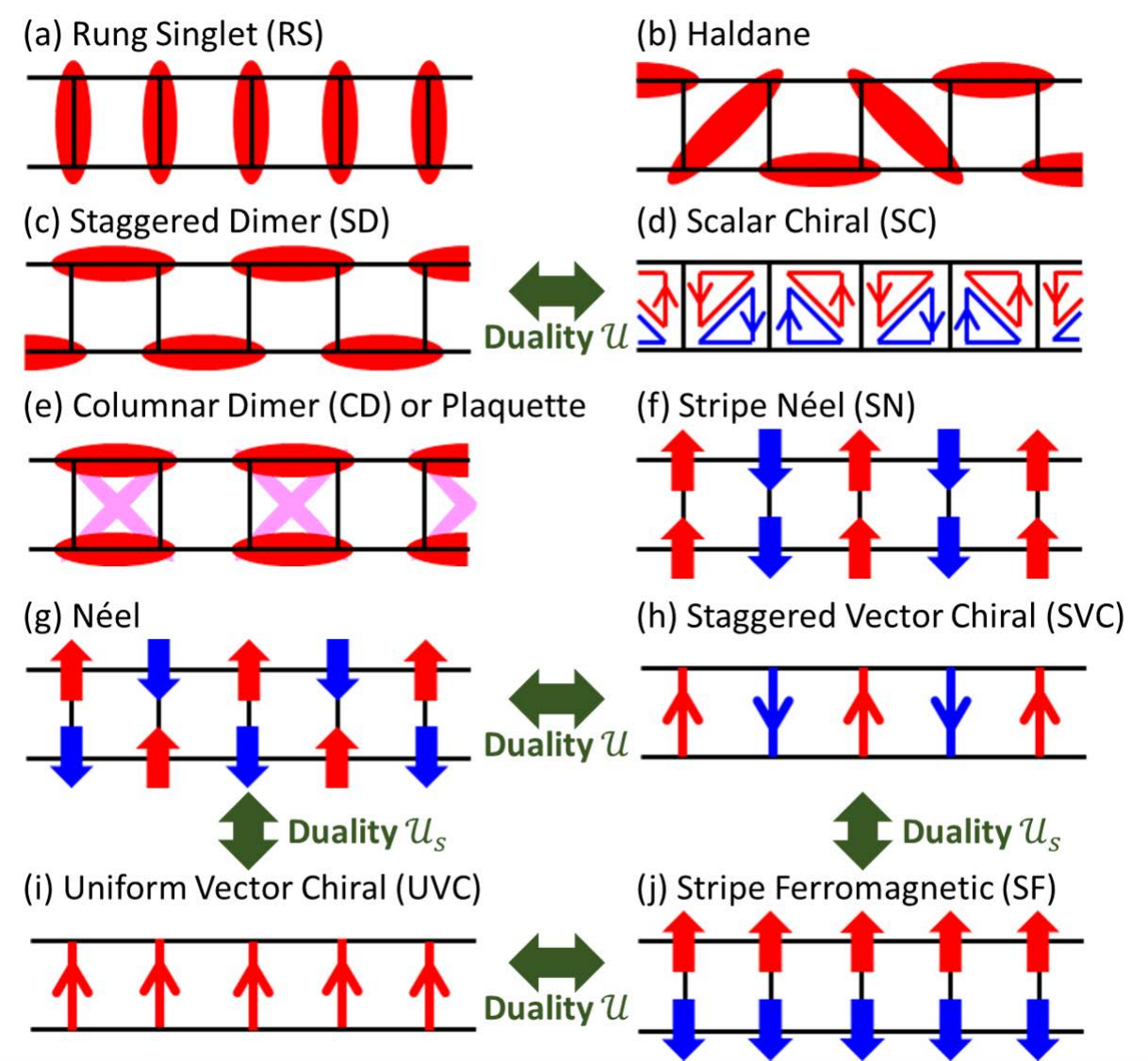}
\caption{\label{fig:ladderstates_chpm}
(Color online) 
Ground states of the ladder models in Fig.\ \ref{fig:laddermodel_chpm}. 
(a,b) Featureless states. 
A red oval indicates a singlet pair $(\ket{\ua\da}-\ket{\da\ua})/\sqrt{2}$. 
The Haldane state is a superposition of various singlet-covering states. 
(c,d,e) Non-magnetic ordered states. 
In the SC state, each directed triangle indicates a positive scalar chirality of three spins along it. 
The CD state, which is predicted by bosonization for weak inter-chain couplings, 
is expected to evolve gradually into a plaquette state for large negative $\Jperp$, 
where two spins on each rung get symmetrized as indicated by the pink diagonal bonds. 
(f,g,h,i,j) Possible short-range spin and VC correlations in the RS [(g,h,i,j)] and Haldane [(f)] states. 
These panels should be understood as typical snapshots, and they may turn into genuine long-range orders 
in the presence of easy-axis anisotropy \cite{Fontaine24XXZ} (see also Appendix \ref{sec:easy-axis}). 
In (f,g,j), each arrow indicates a magnetic moment along a certain axis. 
In (h,i), each directed bond indicates a positive vector chirality along it. 
See Eqs.\ \eqref{eq:O_SD_CD}, \eqref{eq:O_plaq}, \eqref{eq:O_Neel_SN}, \eqref{eq:O_SVC_UVC}, and \eqref{eq:O_SC} 
for the operators that characterize these orders and correlations. 
All the states and the correlations in this figure appear in the phase diagram 
of the $J$-$\Jperp$-$K_{\rm ch}$ ladder \eqref{eq:H_XXX_CCI} with $J>0$ in Fig.\ \ref{fig:phases_chpm}, 
except for (j) the SF correlation which appears for $J<0$ as discussed in Appendix \ref{sec:easy-axis}. 
The duality transformations $\Ucal$ and $\Ucals$ in Eq.\ \eqref{eq:U_Us} relate some of the orders and the correlations, as indicated by green two-headed arrows. 
}
\end{figure}

\begin{figure*}
\includegraphics[width=0.7\textwidth]{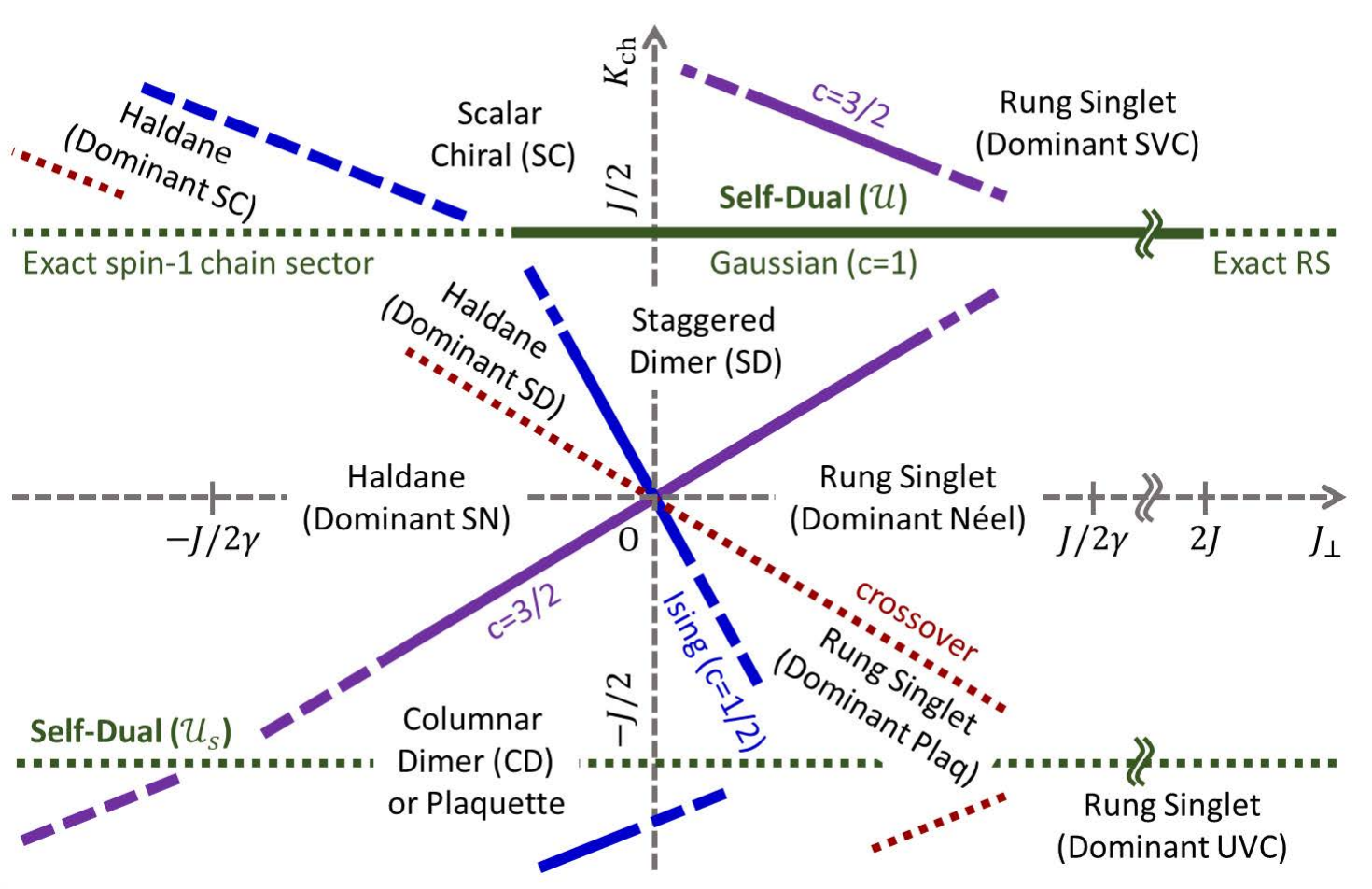}
\caption{\label{fig:phases_chpm}
(Color online) 
Schematic ground-state phase diagram of the $J$-$\Jperp$-$K_{\rm ch}$ ladder \eqref{eq:H_XXX_CCI} with $J>0$, 
predicted by the Abelian bosonization and the duality. 
This diagram is obtained by combining Figs.\ \ref{fig:phases_chpmex}(a,b,c) and setting $\Jdiag=0$. 
The model is self-dual under $\Ucal$ or $\Ucals$ in Eq.\ \eqref{eq:U_Us} at $K_{\rm ch}=J/2$ and $K_{\rm ch}=-J/2$, respectively. 
Along the self-dual line at $\Kch=J/2$, the system has the U(1) symmetry under $\Ucal(\theta)$ in Eq.\ \eqref{eq:U_theta}, 
and the triplon density $n^{\rm B}$ is exactly conserved. 
Along this line, triplons are fully depleted ($n^{\rm B}=0$) for $\Jperp>2J$, leading to the exact RS ground state \cite{Mueller02,Gritsev04,Kolezhuk98}; 
triplons are fully filled ($n^{\rm B}=1$) for $\Jperp\lesssim -0.6 J$, and the ground state resides in the exact spin-$1$ chain sector \cite{Hakovyan08,Fontaine24XXZ}. 
The SD-SC transition occurs on this self-dual line when triplons are partially filled ($0<n^{\rm B}<1$); 
this belongs to the $c=1$ Gaussian universality class, 
according to a field-theoretical analysis around an SU$(4)$ quantum multicritical point \cite{Lecheminant05, Lecheminant06}. 
The phase diagram in the first quadrant ($\Jperp,K_{\rm ch}>0$) as well as its extension to the XXZ model have been studied previously in Ref.\ \cite{Fontaine24XXZ}. 
}
\end{figure*}

% [ Ring exchange on the ladder ]--------------------------------------
The effects of a four-spin exchange have been best understood for the spin-$\frac12$ ladder 
through the applications of powerful field-theoretical \cite{Mueller02, Gritsev04, Totsuka12, Capponi13, Metavitsiadis17} 
and numerical \cite{Laeuchli03, Hikihara03, Hijii02, Hijii03, Mueller02, Capponi13} methods 
as well as an exact spin-chirality duality \cite{Hikihara03, Gritsev04, Momoi03, Hikihara08, Lecheminant05, Lecheminant06, Totsuka12,Hakovyan08}. 
The ground-state phase diagram of an antiferromagnetic ladder with Heisenberg exchanges $J,\Jperp>0$ and a four-spin ring exchange $K>0$ [Fig.\ \ref{fig:laddermodel_chpm}(b)]
includes the rung-singlet (RS) phase, the SD phase, the SC phase, 
and a gapped featureless phase with dominant staggered vector chiral (SVC) correlation (Fig.\ \ref{fig:ladderstates_chpm})
\cite{Laeuchli03, Hikihara03, Mueller02, Gritsev04, Hijii02, Hijii03, Capponi13, Totsuka12, Metavitsiadis17, Furukawa06, Furukawa07}. 
Remarkably, the duality transformation relates the phases for weak and strong $K$. 
Namely, the SC phase is dual to the SD phase, with the SD-SC transition precisely located on the self-dual point $K=J/2$. 
Furthermore, the dominant SVC phase is dual to the RS phase of the antiferromagnetic Heisenberg ladder 
which has dominant N\'eel-type spin correlation; 
in fact, these two phases are continuously connected in the regime of strong $\Jperp>0$ \cite{Lecheminant06} 
(see the first quadrant of Fig.\ \ref{fig:phases_chpm} for an analogous situation). 
Thus, the $J$-$\Jperp$-$K$ ladder in Fig.\ \ref{fig:laddermodel_chpm}(b) showcases an intriguing interplay between spin and chirality degrees of freedom. 

%############################
%############################

%############################
%############################

% [ This paper ]--------------------------------------
In this paper, we aim to explore a spin-chirality interplay further in a {\it simpler} variant of the $J$-$\Jperp$-$K$ ladder. 
Namely, we study the spin-$\frac12$ Heisenberg ladder with a chirality-chirality interaction (CCI) [Fig.\ \ref{fig:laddermodel_chpm}(a)], described by the Hamiltonian 
\begin{equation}\label{eq:H_XXX_CCI}
 H=J \sum_{n=1}^2 \sum_j \Sv_{n,j}\cdot\Sv_{n,j+1}+ \Jperp \sum_j \Sv_{1,j}\cdot\Sv_{2,j}   
 +K_{\rm ch} H_{\rm ch},
\end{equation}
where 
\begin{align}
 H_{\rm ch}=\sum_j 4 \qty( \Sv_{1,j}\times\Sv_{2,j} )\cdot\qty( \Sv_{1,j+1}\times\Sv_{2,j+1} ). \label{eq:H_CCI}
\end{align}
Recently, we have studied the $J$-$\Jperp$-$K_{\rm ch}$ ladder \eqref{eq:H_XXX_CCI} 
with $J, \Jperp,K_{\rm ch}>0$ and weak XXZ anisotropy, 
and have found that the aforementioned phase structure comprising the RS, SD, and SC phases is significantly enriched by the XXZ anisotropy. 
Specifically, near the transitions among these phases, 
two symmetry-protected topological (SPT) phases emerge in an easy-plane regime 
and the N\'eel and SVC ordered phases [Fig.\ \ref{fig:ladderstates_chpm}(g,h)] emerge in an easy-axis regime \cite{Fontaine24XXZ}; 
here, the two phases in each regime are dual to each other. 
In the present work, we focus on the isotropic model \eqref{eq:H_XXX_CCI}, 
and investigate the phase diagram for {\it arbitrary} signs of $\Jperp$ and $K_{\rm ch}$, 
where two types of spin-chirality duality \cite{Hikihara08} represented by $\Ucal$ and $\Ucals$ below can be applied 
to reveal a rich spin-chirality interplay. 
Notably, for negative CCIs $\Kch<0$, 
we find the emergence of a columnar dimer (CD) or plaquette order and uniform vector chiral (UVC) correlation, 
as shown in Fig.\ \ref{fig:phases_chpm}. 

% [ Connections with the ring exchange ]--------------------------------------
The CCI \eqref{eq:H_CCI} can be viewed as part of the four-spin ring exchange 
studied in Refs.\ \cite{Brehmer99, Matsuda00, Nunner02, Schmidt05, Laeuchli03, Hikihara03, Mueller02, Gritsev04, Hijii02, Hijii03, Hikihara08, Totsuka12, Capponi13, Metavitsiadis17, Furukawa06, Furukawa07,Hakovyan08}. 
Indeed, the ring exchange on a plaquette formed by the $j$th and $(j+1)$th rungs [$K$ in Fig.\ \ref{fig:laddermodel_chpm}(b)] is rewritten as 
\begin{equation}\label{eq:P4_S}
\begin{split}
 &P_4(j,j+1)+P_4^{-1}(j,j+1)\\
 &=4\qty( \Sv_{1,j}\times\Sv_{2,j} )\cdot\qty( \Sv_{1,j+1}\times\Sv_{2,j+1} )+P_2(j)P_2(j+1)\\
 &~~~~+\qty( \Sv_{1,j}+\Sv_{2,j} )\cdot\qty( \Sv_{1,j+1}+\Sv_{2,j+1} ), 
\end{split}
\end{equation}
where the first term corresponds to the CCI \eqref{eq:H_CCI}, and 
$P_2(j):=2\Sv_{1,j}\cdot\Sv_{2,j}+1/2$ is the two-spin exchange on the $j$th rung. 
We have found in Ref.\ \cite{Fontaine24XXZ} that the CCI \eqref{eq:H_CCI} brings about a similar phase structure as the ring exchange, 
as shown in the first quadrant of Fig.\ \ref{fig:phases_chpm}. 
Despite the similarities between the CCI \eqref{eq:H_CCI} and the ring exchange \eqref{eq:P4_S}, 
the former has the following advantages in theoretical analyses. 
Firstly, the simple form of the CCI \eqref{eq:H_CCI} in terms of chirality tends to accentuate properties of the chirality-related phases
and to stabilize numerical results (e.g., a weaker dependence on the system size or the bond dimension). 
Secondly, the duality transformations map the CCI \eqref{eq:H_CCI} to two-spin exchanges, 
which leads to a simpler treatment in field theory as described in Sec.\ \ref{sec:EFT_XXX} below. 
This is why we study the $J$-$\Jperp$-$K_{\rm ch}$ ladder \eqref{eq:H_XXX_CCI} rather than the $J$-$\Jperp$-$K$ ladder 
in investigating a spin-chirality interplay in depth. 
We also note that while a four-spin ring exchange naturally has a positive coupling constant $K>0$ \cite{Takahashi77,MacDonald90}, 
a CCI \eqref{eq:H_CCI} that arises as a phonon-mediated interaction in multiferroics \cite{Onoda07, Furukawa08} may have a negative coupling constant $K_{\rm ch}<0$. 

%############################
%############################

% [ Connections with the dimer-dimer interaction ]--------------------------------------
The $J$-$\Jperp$-$K_{\rm ch}$ and $J$-$\Jperp$-$K$ ladders in Fig.\ \ref{fig:laddermodel_chpm} also have close relationship 
with the Heisenberg ladder with an inter-leg dimer-dimer interaction (DDI) studied 
in Refs.\ \cite{Nersesyan97, Pati98, Itoi00, Azaria00, Hijii09, Hijii13, Takayoshi10, Ogino21nvbs,Ogino21spt, ChenJH17,ChenJH19,Robinson19, RenKonik24}. 
The CCI \eqref{eq:H_CCI} can be rewritten as 
\begin{equation}\label{eq:CCI_DDI}
 H_{\rm ch}=4H_{\rm LL}-4H_{\rm DD},
\end{equation}
where 
\begin{subequations}
\begin{align}
H_{\rm LL}=\sum_j \qty(\Sv_{1,j}\cdot\Sv_{1,j+1}) \qty(\Sv_{2,j}\cdot\Sv_{2,j+1}),\\
H_{\rm DD}=\sum_j \qty(\Sv_{1,j}\cdot\Sv_{2,j+1}) \qty(\Sv_{2,j}\cdot\Sv_{1,j+1}).
\end{align}
\end{subequations}
The first term $4H_{\rm LL}$ coincides with the inter-leg DDI. 
When the coupling constant is positive (negative), the inter-leg DDI plays a role of a dimer repulsion (attraction), 
and induces a staggered (columnar) arrangement of singlet dimers, as shown by red ovals in Fig.\ \ref{fig:ladderstates_chpm}(c,e). 
The Heisenberg ladder with an inter-leg DDI thus exhibits an interesting competition 
among the RS and Haldane phases induced by $\Jperp>0$ and $\Jperp<0$, respectively \cite{Strong92,Strong94,Shelton96,Barnes93,White94,Greven96}, 
and the SD and CD phases induced by the DDI \cite{Nersesyan97, Hijii09, Hijii13, Takayoshi10, Ogino21nvbs,Ogino21spt, ChenJH17,ChenJH19,Robinson19, RenKonik24}. 
As we will see, the $J$-$\Jperp$-$K_{\rm ch}$ ladder \eqref{eq:H_XXX_CCI} also exhibits a similar competition among the four phases for relatively weak $K_{\rm ch}$; 
there, field-theoretical analyses \cite{Mueller02, Gritsev04, Metavitsiadis17,Fontaine24XXZ} (see also Appendix \ref{sec:fourspin_bos}) 
suggest that the second term $-4H_{\rm DD}$ in Eq. \eqref{eq:CCI_DDI} only gives a small modification to the effect of the first term $4H_{\rm LL}$. 
However, a crucial difference occurs in the regime of large $|K_{\rm ch}|$, 
where the CCI induces strong SC or VC correlation. 

%There, the inter-leg DDI induces a gapless phase \cite{Pati98, LiYQ98, Itoi00, Azaria00, affleck1986exact, affleck1988critical} in the isotropic case, 
%as opposed to gapped phases induced by a CCI in Fig.\ \ref{fig:phases_ch}. 

%############################
\begin{figure*}
\includegraphics[width=1.0\textwidth]{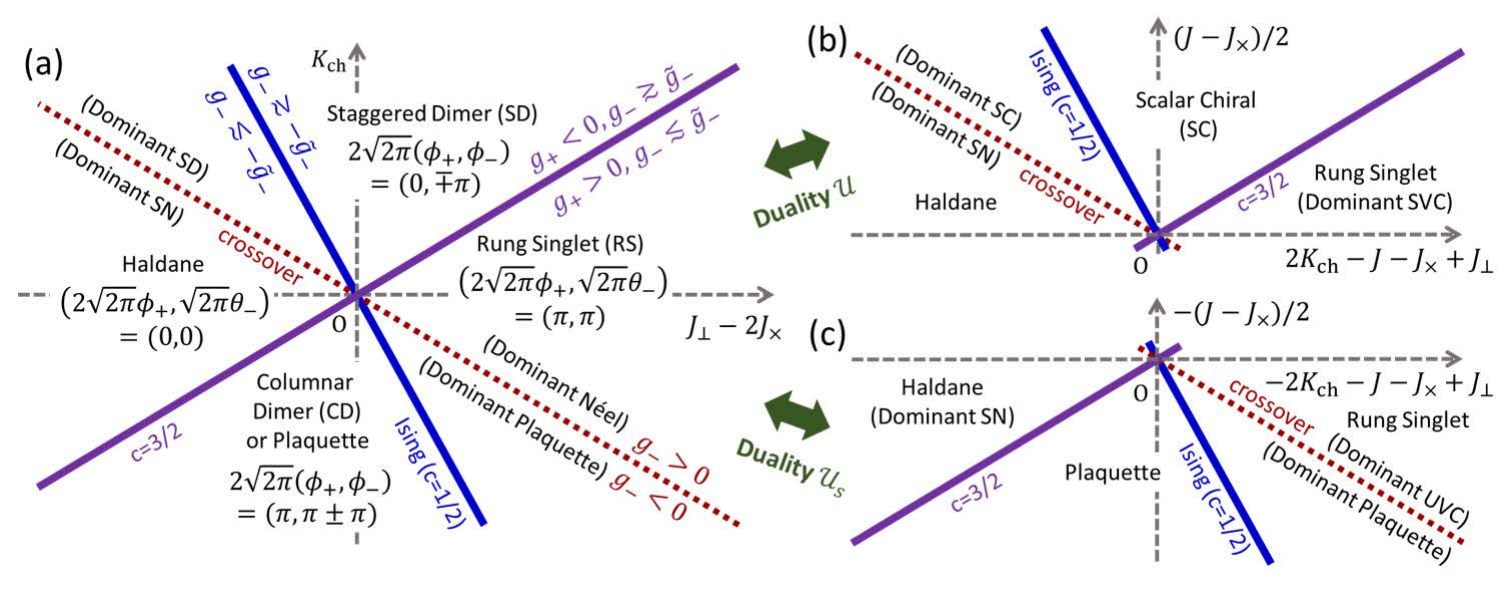}
\caption{\label{fig:phases_chpmex}
(Color online) 
Schematic ground-state phase diagrams of the $J$-$\Jperp$-$\Jdiag$-$K_{\rm ch}$ ladder \eqref{eq:H_XXX_chex}, 
predicted by the Abelian bosonization and the two types of duality. 
(a) Case of weak inter-chain couplings \eqref{eq:H_chpmex_weak}. 
Each phase in (a) is characterized by the locking positions of bosonic fields, as described in Refs.\ \cite{Takayoshi10,Ogino21nvbs,Ogino21spt}. 
(b,c) Dual counterparts to (a) obtained by the transformations $\Ucal$ and $\Ucals$, respectively, in Eqs.\ \eqref{eq:U_Us}; 
see Eq.\ \eqref{eq:H_chpmex_duality} for the transformation of the coupling constants. 
These phase diagrams are valid under the condition \eqref{eq:condition_dual_chpmex}. 
The transition and crossover lines in (a) are given by Eqs.\ \eqref{eq:line_WZW}, \eqref{eq:line_Ising}, and \eqref{eq:line_crossover} 
while those in (b,c) are given by Eqs.\ \eqref{eq:line_WZW_dual}, \eqref{eq:line_Ising_dual}, and \eqref{eq:line_crossover_dual}. 
}
\end{figure*}
%############################

% [ Main results ]--------------------------------------
Our main results for the $J$-$\Jperp$-$K_{\rm ch}$ ladder with $J>0$ and arbitrary signs of $\Jperp$ and $K_{\rm ch}$ 
are shown in the schematic phase diagram in Fig.\ \ref{fig:phases_chpm}. 
This phase diagram is based on a field-theoretical analysis for weak inter-chain couplings as well as the two types of spin-chirality duality. 
The phase diagram consists of the RS and Haldane phases as well as the three non-magnetic ordered phases (SD, SC, and CD) described above. 
The RS and Haldane phases are {\it featureless} in the sense that the bulk ground state is unique and does not spontaneously break any symmetry. 
Nonetheless, these phases can be divided into several regions with different dominant correlations in terms of spin, dimer, or chirality. 
We perform numerical analyses based on 
infinite density-matrix renormalization group (iDMRG) \cite{McCulloch07,McCulloch08, itensor,itensor-r0.3} 
and exact diagonalization \cite{QuSpin17, QuSpin19} 
to confirm the predicted phase structure and critical properties. 

In various respects, the analyses conducted in this work are extensions of those in our previous work \cite{Fontaine24XXZ}. 
An analysis of the regime of strong CCIs $\Kch> J/2$ by combining the Abelian bosonization with the spin-chirality duality (under $\Ucal$) 
has already been presented in Ref.\ \cite{Fontaine24XXZ}. 
Importantly, in the present work, 
we also combine the bosonization with the second duality (under $\Ucals$) to investigate the regime of strong negative CCIs $\Kch< -J/2$. 
Additionally, we employ the fermionization to analyze correlation functions of different operators in the RS and Haldane phases. 
Numerical methods for identifying phase boundaries and critical properties are also similar to those in Ref.\ \cite{Fontaine24XXZ}. 
In the present work, we also calculate structure factors and correlation functions to investigate changes in dominant correlation. 

% [ Organization of the paper ]--------------------------------------
The rest of this paper is organized as follows. 
In Sec.\ \ref{sec:EFT_XXX}, we analyze the $J$-$\Jperp$-$K_{\rm ch}$ ladder \eqref{eq:H_XXX_CCI} 
by means of the Abelian bosonization and the spin-chirality duality, 
and derive the schematic phase diagram in Fig.\ \ref{fig:phases_chpm}. 
In Sec.\ \ref{sec:numer}, we present numerical analyses. 
In Sec.\ \ref{sec:summary}, we present a summary of the present study and an outlook for future studies. 
In Appendix \ref{sec:easy-axis}, we analyze an XXZ variant of the model \eqref{eq:H_XXX_CCI} in the easy-axis regime, 
which can provide a simple picture for the emergence of N\'eel, SVC, and UVC correlations. 
In Appendix \ref{sec:fourspin_bos}, we provide field-theoretical expressions of the four-spin interactions in Eq.\ \eqref{eq:CCI_DDI}, 
which are used in Sec.\ \ref{sec:EFT_XXX}. 

%%%%%%%%%%%%%%%%%%%%%%%%%%%%%%%%%%%%%%%%%%%%%%%%
\section{Field-Theoretical Analysis}\label{sec:EFT_XXX}
%%%%%%%%%%%%%%%%%%%%%%%%%%%%%%%%%%%%%%%%%%%%%%%%

% [ Extended XXX-CCI ladder ]--------------------------------------
In this section, we analyze the ground-state phase diagram of the $J$-$\Jperp$-$K_{\rm ch}$ ladder \eqref{eq:H_XXX_CCI} 
by applying the Abelian bosonization and the two types of spin-chirality duality. 
To this end, following our previous work \cite{Fontaine24XXZ}, we slightly extend the model \eqref{eq:H_XXX_CCI} by adding diagonal exchange couplings [Fig.\ \ref{fig:laddermodel_chpm}(a)] as 
\begin{equation}\label{eq:H_XXX_chex}
 H_\text{ex}=H+\Jdiag \sum_j \qty( \Sv_{1,j}\cdot\Sv_{2,j+1} + \Sv_{2,j}\cdot\Sv_{1,j+1} ).
\end{equation}
Although we set $\Jdiag=0$ at the end of the analysis, the roles of the duality transformations become more transparent in this extended model. 
The case of $\Jperp-2\Jdiag,K_{\rm ch}>0$ has been analyzed in Ref.\ \cite{Fontaine24XXZ}, 
and below we extend it for arbitrary signs of these couplings. 

% [ Our strategy ]--------------------------------------
We start by analyzing the regime 
\begin{equation}\label{eq:H_chpmex_weak}
 |\Jperp|, |\Jdiag|, 2|K_{\rm ch}| \ll J,
\end{equation}
where the model \eqref{eq:H_XXX_chex} can be viewed as weakly coupled Heisenberg chains. 
In this regime, we can apply the Abelian bosonization \cite{Giamarchi04,Gogolin04} to formulate an effective field theory,  
as have been done for a variety of ladder models 
\cite{Strong92,Strong94,Shelton96, Nersesyan97,Mueller02,Takayoshi10,ChenJH17,ChenJH19,Robinson19, RenKonik24, 
Ogino21nvbs,Ogino21spt,Fontaine24XXZ,
Kim00, Vekua06, Hikihara10,Ogino22,
Nakamura03, FuruyaSato21, Fuji15, Fuji16, Furukawa08, Furukawa12}. 
Below we take notations similar to those in Refs.\ \cite{Takayoshi10,Ogino21nvbs,Ogino21spt,Ogino22,Fontaine24XXZ} 
so that we can simplify the explanation of the formalism. 
After analyzing the regime of Eq.\ \eqref{eq:H_chpmex_weak}, we apply the two types of spin-chirality duality to discuss the regime of large $|K_{\rm ch}|$. 

%************************************************
\subsection{Abelian bosonization for weak inter-chain couplings}\label{sec:bos}
%************************************************

% [ Effective H for coupled chains ]--------------------------------------
As described in Refs.\ \cite{Takayoshi10,Ogino21nvbs,Ogino21spt,Ogino22,Fontaine24XXZ}, 
the low-energy effective Hamiltonian of the model \eqref{eq:H_XXX_chex} for weak inter-chain couplings \eqref{eq:H_chpmex_weak} 
is given by perturbed two-component Tomonaga-Luttinger liquids (TLLs) 
\begin{equation}\label{eq:H_chex_eff}
\begin{split}
H_{\rm ex}^{\rm eff}
&=\int \dd x \sum_{\nu=\pm} \frac{v_\nu}{2} \qty[ \frac1{K_\nu} \qty(\partial_x\phi_\nu) ^2+ K_\nu \qty(\partial_x\theta_\nu)^2] \\
&+ g_+ \cos(2\sqrt{2\pi}\phi_+) + g_- \cos(2\sqrt{2\pi}\phi_-) \\
&+ \tilde{g}_- \cos(\sqrt{2\pi}\theta_-)-\gamma_1 \cos(2\sqrt{2\pi}\phi_+) \cos(\sqrt{2\pi}\theta_-)\\
&+\gamma_{\rm bs} \cos(2\sqrt{2\pi}\phi_+) \cos(2\sqrt{2\pi}\phi_-)+\dots, 
\end{split}
\end{equation}
where $(\phi_+,\theta_+)$ and $(\phi_-,\theta_-)$ are dual pairs of bosonic fields in the symmetric ($+$) and antisymmetric ($-$) channels, respectively, 
and ellipses indicate terms that have higher scaling dimensions. 
The bare coupling constants are given by  
\begin{subequations}\label{eq:coeff_cos}
\begin{align}
 g_\pm &= \frac{1}{2a} \qty[ \ab^2 \qty(\Jperp- 2\Jdiag) \mp 24\db^2 K_{\rm ch} ], \label{eq:coeff_cos_gpm}\\
 \tilde{g}_-&=\frac{\ab^2}{a}  (\Jperp-2\Jdiag), \label{eq:coeff_gtm}\\
 \gamma_1&=\frac{b_1^2}{2a} (\Jperp+2\Jdiag-C\Kch), \label{eq:coeff_gamma1}
\end{align}
\end{subequations}
and $\gamma_{\rm bs}$ arises from the backscattering in each chain as well as the CCI. 
Here, $\ab$, $b_1$, and $\db$ are coefficients in the bosonized expressions of spin and dimer operators \cite{Takayoshi10}, 
and $C>0$ is defined in Appendix \ref{sec:fourspin_bos}. 
The velocities $v_\pm$ and the TLL parameters $K_\pm$ in the $\pm$ channels 
are in general modified from $v=\pi Ja/2$ and $K=1/2$ in the decoupled Heisenberg chains by the effects of the inter-chain couplings. 

%************************************************
\subsection{Phase diagram from bosonization}\label{sec:bos_phases}
%************************************************

% [ Symmetric and antisymmetric channels ]--------------------------------------
In the effective Hamiltonian \eqref{eq:H_chex_eff}, the strongly relevant $g_\pm$ and $\tilde{g}_-$ terms
(with the scaling dimensions $2K_\pm$ and $(2K_-)^{-1}$, respectively) 
are expected to determine the phase structure dominantly. 
We tentatively neglect the $\gamma_1$ and $\gamma_{\rm bs}$ terms, 
which are marginal in the limit of the decoupled Heisenberg chains. 
Then, the $\pm$ channels are mutually decoupled in the effective Hamiltonian \eqref{eq:H_chex_eff}, 
and the phase structure can be analyzed in the same way as in Refs.\ \cite{Takayoshi10, Ogino21nvbs,Ogino21spt,Ogino22,Fontaine24XXZ}. 
Namely, in the $+$ channel described by the sine-Gordon model, 
the strongly relevant $g_+$ term locks $\phi_+$ at distinct positions depending on the sign of $g_+$; 
a $c=1$ Gaussian transition occurs at $g_+=0$. 
In the $-$ channel described by the dual-field double sine-Gordon model, 
the strongly relevant $g_- $ and $\tilde{g}_-$ terms with close scaling dimensions crucially compete. 
In this model, $|g_-|\gtrsim |\tilde{g}_-|$ ($|g_-|\lesssim |\tilde{g}_-|$) leads to the locking of $\phi_-$ ($\theta_-$), 
and an $c=1/2$ Ising transition occurs at $|g_-|\approx |\tilde{g}_-|$ \cite{Shelton96,LECHEMINANT2002502}. 
We will address the effects of the marginal $\gamma_1$ and $\gamma_{\rm bs}$ terms at the end of this section. 

% [ Schematic phase diagram ]--------------------------------------
Based on the above analysis and using the coupling constants in Eq.\ \eqref{eq:coeff_cos}, 
we obtain the schematic phase diagram in Fig.\ \ref{fig:phases_chpmex}(a). 
In this figure, there are four phases, each characterized by the dominance of $\Jperp-2\Jdiag$ or $K_{\rm ch}$ with either a positive or negative sign. 
Along the purple line with 
\begin{equation}\label{eq:line_WZW}
  K_{\rm ch} = \gamma \qty(\Jperp- 2\Jdiag) \ne 0,
\end{equation}
Gaussian and Ising transitions simultaneously occur in the $+$ and $-$ channels, respectively. 
Here, the proportionality constant $\gamma$ is estimated to be $\gamma= \ab^2/(24\db^2)\approx 0.75 $ \cite{Takayoshi10}. 
This simultaneous occurrence of the transitions in the two channels is due to the SU(2) symmetry of the original model \eqref{eq:H_XXX_CCI}, 
and is described by the SU(2)$_2$ Wess-Zumino-Witten (WZW) theory with the central charge $c=3/2$ 
\cite{Hijii02, Hijii03, Mueller02, Capponi13, Gritsev04, Nersesyan97, Hijii09, Hijii13, ChenJH17,ChenJH19,Robinson19, RenKonik24, Mueller02, Tsuchiizu02}. 
Meanwhile, along the blue line with 
\begin{equation}\label{eq:line_Ising}
  K_{\rm ch} = -3\gamma \qty(\Jperp- 2\Jdiag)\ne 0,
\end{equation}
an Ising transition occurs in the $-$ channel while the $+$ channel remains gapped. 
We note that a phase structure similar to Fig.\ \ref{fig:phases_chpmex}(a) has also been obtained for a Heisenberg ladder with an inter-leg DDI 
\cite{Takayoshi10,ChenJH17,ChenJH19,Robinson19, RenKonik24}. 

% [ SD and CD phases ]--------------------------------------
In Fig.\ \ref{fig:phases_chpmex}(a), the phases induced by a dominant positive or negative CCI $K_{\rm ch}$ 
are characterized by the spontaneous formation of staggered and columnar singlet dimer patterns, respectively, 
as shown by red ovals in Fig.\ \ref{fig:laddermodel_chpm}(c,e). 
These can be interpreted as a consequence of a dimer repulsion or attraction induced by $K_{\rm ch}>0$ and $K_{\rm ch}<0$, respectively; see Eq.\ \eqref{eq:CCI_DDI}. 
These dimer patterns are characterized by the SD and CD order parameters
\begin{subequations}\label{eq:O_SD_CD}
\begin{align}
\expval{\mathcal{O}_{\text{SD}}(j)} &= \frac14 
\langle \bm{S}_{1,j-1}\cdot\bm{S}_{1,j} - \bm{S}_{2,j-1}\cdot\bm{S}_{2,j} \notag\\
&~~~~~~-\bm{S}_{1,j}\cdot\bm{S}_{1,j+1} + \bm{S}_{2,j}\cdot\bm{S}_{2,j+1} \rangle, \label{eq:O_SD}\\
\expval{\mathcal{O}_{\text{CD}}(j)} &= \frac14 
\langle \bm{S}_{1,j-1}\cdot\bm{S}_{1,j} + \bm{S}_{2,j-1}\cdot\bm{S}_{2,j} \notag\\
&~~~~~~-\bm{S}_{1,j}\cdot\bm{S}_{1,j+1} - \bm{S}_{2,j}\cdot\bm{S}_{2,j+1} \rangle, \label{eq:O_CD}
\end{align}
\end{subequations}
which are bosonized as 
\begin{subequations}\label{eq:O_SD_CD_bos}
\begin{align}
 \mathcal{O}_{\text{SD}}(j) 
 &=-(-1)^j (3\bar{d}) \cos \qty(\sqrt{2\pi}\phi_+) \sin \qty(\sqrt{2\pi}\phi_-), \label{eq:O_SD_bos} \\
 \mathcal{O}_{\text{CD}}(j) 
 &=-(-1)^j (3\bar{d}) \sin \qty(\sqrt{2\pi}\phi_+) \cos \qty(\sqrt{2\pi}\phi_-). \label{eq:O_CD_bos}
\end{align}
\end{subequations}
We can confirm that these order parameters acquire finite values for the locking positions in Fig.\ \ref{fig:phases_chpmex}(a). 
For example, in the CD phase with $2\sqrt{2\pi}(\phi_+,\phi_-)=(\pi,\pi\pm\pi)$, 
we have $\expval{\mathcal{O}_{\text{CD}}(j)}=\pm(-1)^j c_\text{CD}$, where $c_\text{CD}$ is a constant independent of $j$. 

% [ Plaquette ]--------------------------------------
However, the CD order may not give an appropriate picture over the entire phase, 
given the fact that the same phase continues over the self-dual line at $K_{\rm ch}=-J/2$ and further for large negative $\Jperp$ and $K_{\rm ch}$, 
as shown in Fig.\ \ref{fig:phases_chpm}. 
In particular, the CD order parameter \eqref{eq:O_CD} is {\it not} invariant under the duality transformation $\Ucals$ defined later in Eq.\ \eqref{eq:U_Us} 
while this transformation does not change the broken symmetry of the concerned phase.  
Furthermore, the two spins on each rung are expected to be gradually symmetrized for larger ferromagnetic $\Jperp<0$, 
as depicted by the diagonal singlet formation in Fig.\ \ref{fig:laddermodel_chpm}(e).
We therefore modify the CD order parameter \eqref{eq:O_CD} slightly to introduce the plaquette order parameter 
\begin{equation}\label{eq:O_plaq}
\begin{split}
\expval{\mathcal{O}_{\text{plaq}}(j)} &= \frac14 
\langle \qty(\bm{S}_{1,j-1}+\bm{S}_{2,j-1}) \cdot \qty(\bm{S}_{1,j} +\bm{S}_{2,j}) \\
&~~~~~~- \qty(\bm{S}_{1,j} + \cdot\bm{S}_{2,j})\cdot \qty(\bm{S}_{1,j+1}+\bm{S}_{2,j+1})\rangle. 
\end{split}
\end{equation}
This operator is symmetric for two spins on any rung, and also invariant under $\Ucals$. 
Below we shall call the concerned phase the plaquette phase, 
and this gives a persistent picture of the phase up to large negative $\Jperp$ and $K_{\rm ch}$. 

% [ RS and Haldane phases ]--------------------------------------
In Fig.\ \ref{fig:phases_chpmex}(a), the phases induced by dominant positive or negative $\Jperp-2\Jdiag$ 
are characterized as the RS and Haldane phases, respectively, 
which are known in the studies of Heisenberg ladders \cite{Strong92,Strong94,Shelton96,Barnes93,White94,Greven96}. 
Both of these phases are gapped and featureless, i.e., they have a unique ground state below an excitation gap, and do not spontaneously break any symmetry. 
Despite this featureless nature, 
each of these phases can be divided into two regions in terms of dominant short-range correlation. 
To discuss such correlation, in addition to the SD and CD operators \eqref{eq:O_SD_CD}, 
we introduce the N\'eel and stripe N\'eel (SN) operators along the $z$ axis [see Fig.\ \ref{fig:ladderstates_chpm}(f,g)]
\begin{subequations}\label{eq:O_Neel_SN}
\begin{align}
\mathcal{O}_{\text{N\'eel}}(j) &= \frac14 \qty( S^z_{1,j} - S^z_{2,j} - S^z_{1,j+1} + S^z_{2,j+1} ), \label{eq:O_Neel}\\
\mathcal{O}_{\text{SN}}(j) &= \frac14 \qty( S^z_{1,j} + S^z_{2,j} - S^z_{1,j+1} - S^z_{2,j+1}), \label{eq:O_SN}
\end{align}
\end{subequations}
which are bosonized as 
\begin{subequations}\label{eq:O_Neel_SN_bos}
\begin{align}
 \mathcal{O}_{\text{N\'eel}}(j)
 &= -(-1)^j \bar{a} \sin \qty(\sqrt{2\pi}\phi_+) \sin \qty(\sqrt{2\pi}\phi_-), \label{eq:O_Neel_bos}\\
 \mathcal{O}_{\text{SN}}(j)
 &=(-1)^j \bar{a} \cos \qty(\sqrt{2\pi}\phi_+) \cos \qty(\sqrt{2\pi}\phi_-). \label{eq:O_SN_bos}
\end{align}
\end{subequations}
In the RS phase, $\sin \qty(\sqrt{2\pi}\phi_+)$ in the CD and N\'eel operators 
in Eqs.\ \eqref{eq:O_CD_bos} and \eqref{eq:O_Neel_bos} acquires a finite average 
while $\cos \qty(\sqrt{2\pi}\phi_-)$ and $\sin \qty(\sqrt{2\pi}\phi_-)$ in these operators fluctuate significantly 
owing to the locking of $\theta_-$; see the field locking position in Fig.\ \ref{fig:phases_chpmex}(a). 
Despite such fluctuations, when $g_-<0$, the $g_-$ term in Eq.\ \eqref{eq:H_chex_eff} 
favors states around $\sqrt{2\pi}\phi_-=0$ or $\pi$ over short distances, 
leading to short-range CD correlation. 
Likewise, $g_->0$ leads to short-range N\'eel correlation. 
A similar argument also applies to the Haldane phase, 
which is then divided into the regions with dominant SD or SN correlation.
In these ways, a change in dominant correlation occurs at $g_-=0$, i.e., 
\begin{equation}\label{eq:line_crossover}
 K_{\rm ch}=-\gamma (\Jperp-2\Jdiag) \ne 0, 
\end{equation}
which is shown by the red dotted line in Fig.\ \ref{fig:phases_chpmex}(a). 

% [ Fermionized theory ]--------------------------------------
The above argument is also supported by the fermionized field theory 
\cite{Gogolin04,Shelton96,Nersesyan97,Mueller02,Gritsev04,ChenJH17,ChenJH19,Robinson19, RenKonik24}. 
Via the fermionization, the two-component TLLs at $K=1/2$ perturbed by the strongly relevant $g_\pm$ and $\gt$ terms in Eq.\ \eqref{eq:H_chex_eff} 
can be mapped to four-component free Majorana fields (i.e., noncritical two-dimensional Ising models) 
with a triplet of mass $m_{\rm t}$ and a singlet of mass $m_{\rm s}$, where
\begin{subequations}\label{eq:mtms}
\begin{align}
 m_{\rm t} &= \pi a g_+ = \pi a (\gt_- - g_-) \notag\\
 &= 12\pi \db^2 \qty[\gamma \qty(\Jperp- 2\Jdiag) - K_{\rm ch} ], \\
 m_{\rm s} &= - \pi a (\gt_- + g_-) \notag\\
 &=-12\pi \db^2 \qty[3\gamma \qty(\Jperp- 2\Jdiag) + K_{\rm ch} ] .
\end{align}
\end{subequations}
A sign change in $m_{\rm t}$ or $m_{\rm s}$ corresponds to the $c=3/2$ and $c=1/2$ transitions in Eqs.\ \eqref{eq:line_WZW} and \eqref{eq:line_Ising}, respectively. 
Furthermore, a change in dominant correlation inside the RS and Haldane phases occurs at $m_{\rm t}=-m_{\rm s}$ \cite{ChenJH17}, 
which corresponds to Eq.\ \eqref{eq:line_crossover}. 
To see this, it is convenient to express the dimer and spin operators \eqref{eq:O_SD_CD_bos} and \eqref{eq:O_Neel_SN_bos} as
\begin{subequations}\label{eq:O_SD_CD_Neel_SN_fer}
\begin{align}
 \mathcal{O}_{\text{SD}}(j) 
 &\sim (-1)^j \sigma_1\sigma_2\sigma_3\mu, \\
 \mathcal{O}_{\text{CD}}(j) 
 &\sim (-1)^j \mu_1\mu_2\mu_3\sigma, \\
 \mathcal{O}_{\text{N\'eel}}(j)
 &\sim (-1)^j \mu_1\mu_2\sigma_3\mu , \\
 \mathcal{O}_{\text{SN}}(j)
 &\sim (-1)^j \sigma_1\sigma_2\mu_3\sigma,
\end{align}
\end{subequations}
where $\sigma_\alpha$ and $\mu_\alpha~(\alpha=1,2,3)$ are the order and disorder fields of the ``triplet'' Ising models, 
and $\sigma$ and $\mu$ are the similar fields for the ``singlet'' Ising model. 
In the RS phase with $m_{\rm t}>0$ and $m_{\rm s}<0$, the disorder fields acquire finite averages $\langle \mu_\alpha \rangle\ne 0$ and $\langle \mu \rangle \ne 0$ 
while the order fields show decaying correlations 
\begin{subequations}\label{eq:sigma_corr_decay}
\begin{align}
 \langle \sigma_\alpha (0) \sigma_\alpha (r) \rangle &\sim K_0 \qty(\frac{m_{\rm t}r}{v}) ~(\alpha=1,2,3) ,\label{eq:corr_sigmat}\\
 \langle \sigma (0) \sigma (r) \rangle &\sim K_0 \qty(\frac{|m_{\rm s}|r}{v}) ,\label{eq:corr_sigmas}
\end{align}
\end{subequations}
where $v$ is the velocity. 
Here, $K_0(\cdot)$ is the MacDonald function, which behaves at long distances as 
\begin{equation}
 K_0(r) \approx  \sqrt{\frac{\pi}{2r}} e^{-r}~~(r\gg 1). 
\end{equation}
Therefore, in the RS phase, the behaviors of the N\'eel and CD correlations are mainly given by Eqs.\ \eqref{eq:corr_sigmat} and \eqref{eq:corr_sigmas}, respectively, 
and $m_{\rm t}<|m_s|$ ($m_{\rm t}>|m_s|$) leads to the dominance of the N\'eel (CD) correlation. 
A similar argument applies to the Haldane phase (with $m_{\rm t}<0$ and $m_{\rm s}>0$) by exchanging the roles of the order and disorder fields. 
Namely, $|m_{\rm t}|<m_s$ ($|m_{\rm t}|>m_s$) leads to the dominance of the SN (SD) correlation, 
whose decaying behavior is given mainly by the correlation of $\mu_3$ ($\mu$). 

\begin{figure*}
\includegraphics[width=0.7\textwidth]{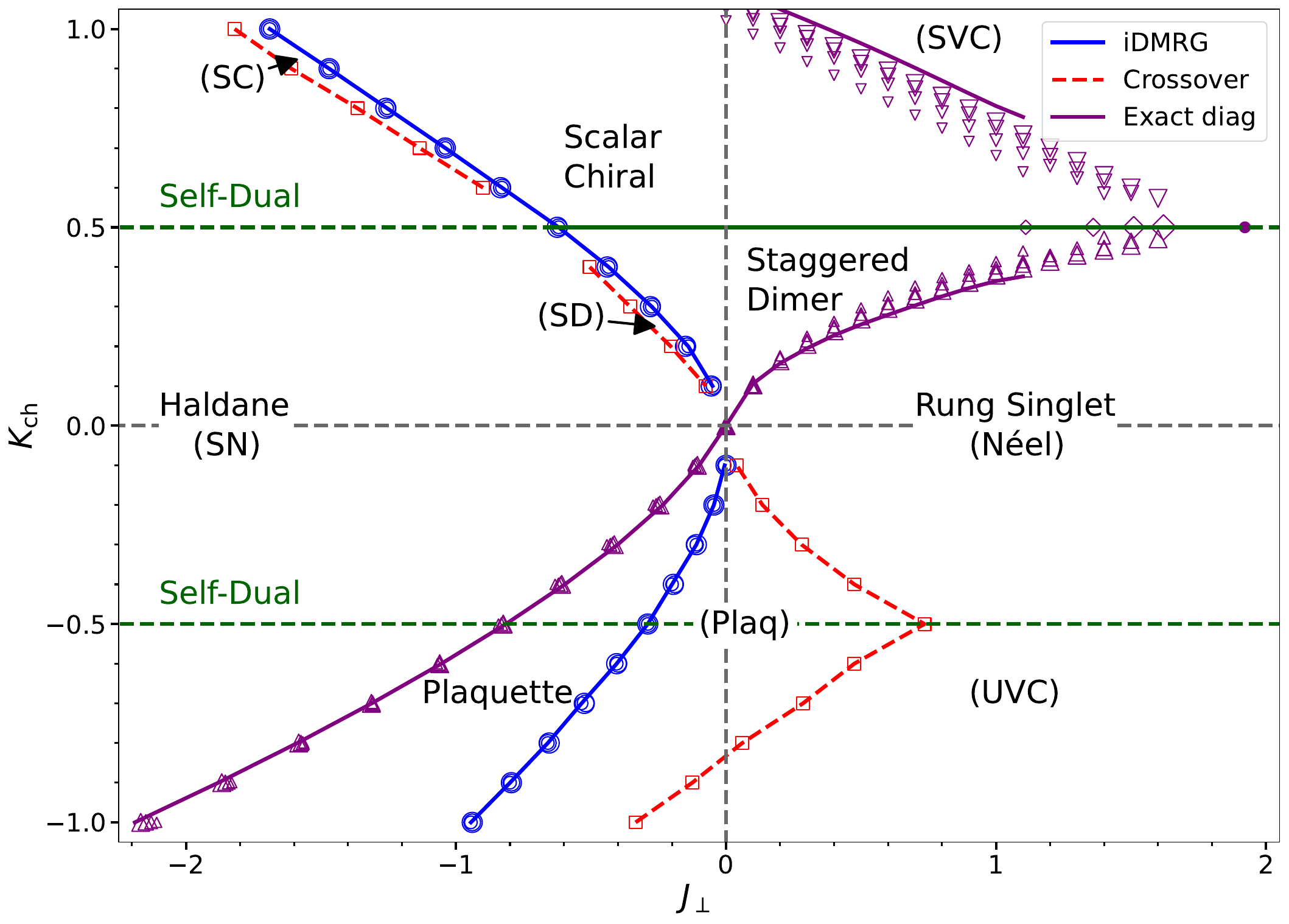}
\caption{\label{fig:pd_iDMRG_Exact}
(Color online) 
Ground-state phase diagram of the $J$-$\Jperp$-$\Kch$ ladder \eqref{eq:H_XXX_CCI} with $J=1$, obtained numerically. 
Ising transition points (blue circles) are determined by iDMRG from the (local) maxima of the correlation length.  
Small, medium, and large circles correspond to the bond dimensions $\chi=128$, $192$, and $256$, respectively, 
and large circles are interpolated by blue solid lines as a guide for the eye. 
Purple triangles and diamonds indicate the sign change of the polarization amplitude $z_{\rm rung}$ in Eq.\ \eqref{eq:z_rung} calculated by exact diagonalization. 
For $\Kch\ne 0$ and $J/2$, these points are expected to correspond to the $c=3/2$ transitions in Fig.\ \ref{fig:phases_chpm}. 
Symbols with increasing marker sizes correspond to the ladder lengths $L = 8$, $10$, $12$, and $14$. 
For $K_{\rm ch} \ge 0$, up- and down-pointing triangles indicate points 
at which $z_{\rm rung}$ becomes negative and positive, respectively, with increasing $K_{\rm ch}$; 
these data are taken from Ref.\ \cite{Fontaine24XXZ}. 
For $K_{\rm ch} = J/2$ and for $K_{\rm ch}<0$, the sign change of $z_{\rm rung}$ is investigated along the $\Jperp$ axis, 
and the zeros of $z_{\rm rung}$ are shown by diamonds and up-pointing triangles, respectively. 
For $\Jperp\le 1.0$, where the $L$-dependence of the points is relatively smooth, 
extrapolation of the transition points to the infinite-size limit is conducted, and the results are interpolated by purple solid lines. 
Red squares interpolated by dashed lines indicate changes in dominant correlation inside the RS and Haldane phases. 
These points are determined by iDMRG with $\chi = 128$ as described in Sec.\ \ref{sec:DC}. 
The types of dominant correlation are indicated in parentheses. 
}
\end{figure*}

% [ Effects of marginal interactions ]--------------------------------------
Finally, we briefly discuss the effects of the marginal $\gamma_1$ and $\gamma_{\rm bs}$ terms in the effective Hamiltonian \eqref{eq:H_chex_eff}, 
which have been neglected so far. 
The $\gamma_{\rm bs}$ term is present even in the limit of decoupled chains, and is negative in this limit; 
away from this limit, $\gamma_{\rm bs}$ also has a contribution $C'\Kch$ with $C'>0$ from the CCI 
[see the $O_a$ term in Eq.\ \eqref{eq:HLL_bos} in Appendix \ref{sec:fourspin_bos}]. 
The field locking positions in the SD and plaquette phases in Fig.\ \ref{fig:phases_chpmex}(a) are disfavored by $\gamma_{\rm bs}<0$. 
Meanwhile, when $\gamma_1>0$ ($\gamma_1<0$), 
the $\gamma_1$ term favors (disfavors) both the field locking positions of the RS and Haldane phases. 
Therefore, in the fourth quadrant of Fig.\ \ref{fig:phases_chpmex}(a) (with $\Jdiag$ set to zero), 
where $\gamma_1$ in Eq.\ \eqref{eq:coeff_gamma1} and $\gamma_{\rm bs}$ are particularly enhanced with positive and negative signs, respectively, 
the plaquette-RS phase boundary is expected to be significantly influenced by the marginal terms 
in such a way that the RS phase expands as one deviates from the origin. 
In fact, in the numerically obtained phase diagram shown later in Fig.\ \ref{fig:pd_iDMRG_Exact}, 
the plaquette-RS boundary is located entirely on the $\Jperp<0$ side, which can be interpreted from this effect. 
We note that the contribution of the CCI to $\gamma_1$ in Eq.\ \eqref{eq:coeff_gamma1} 
occurs only through  the second term $-4H_{\rm DD}$ in Eq.\ \eqref{eq:CCI_DDI}; 
see the $O_b$ term in Eq.\ \eqref{eq:HDD_bos} in Appendix \ref{sec:fourspin_bos}.  
Therefore, the effect of the marginal $\gamma_1$ term on the plaquette-RS boundary is expected to be weaker 
in the Heisenberg ladder with an inter-leg DDI studied 
in Refs.\ \cite{Nersesyan97, Hijii09, Hijii13, Takayoshi10, Ogino21nvbs,Ogino21spt, ChenJH17,ChenJH19,Robinson19, RenKonik24}. 
This qualitatively explains why the plaquette-RS boundary is located entirely on the $J_\perp > 0$ side 
in the numerical results in Refs.\ \cite{Hijii13,ChenJH19,Robinson19}. 
To discuss the effects of the marginal perturbations more seriously, 
one must solve renormalization group flow equations as in Ref.\ \cite{Metavitsiadis17}, which is beyond the scope of this paper. 

%************************************************
\subsection{Spin-chirality duality}\label{sec:duality}
%************************************************

% [ Duality transformations ]--------------------------------------
The two types of spin-chirality duality 
\cite{Hikihara03, Gritsev04, Momoi03, Hikihara08, Lecheminant05, Lecheminant06, Totsuka12,Hakovyan08} 
can be described by the unitary transformations
\begin{equation}\label{eq:U_Us}
 \Ucal:=\Ucal\qty(\frac{\pi}2),~~\Ucals:=\Ucals\qty(\frac{\pi}2),
\end{equation}
where 
\begin{subequations}
\begin{align}
 \Ucal (\theta) &= \exp(i\theta \sum_j n^{\rm B}_j) \label{eq:U_theta},\\
 \Ucals (\theta) &= \exp[i\theta \sum_j (-1)^j n^{\rm B}_j]~~(\theta\in\mathbb{R}) \label{eq:Us_theta}
\end{align}
\end{subequations}
with $n^{\rm B}_j:=\Sv_{1,j}\cdot\Sv_{2,j}+3/4$ being the projection operator onto the triplet states on the $j$th rung 
(i.e., a local occupation number of a triplon). 
The duality under $\Ucal$ was originally introduced by Hikihara {\it et al.}\ \cite{Hikihara03} 
to discuss the phase diagram of the $J$-$\Jperp$-$K$ ladder in Fig.\ \ref{fig:laddermodel_chpm}(b) with $J,K>0$. 
Later, the duality under $\Ucals$ was introduced by Hikihara and Yamamoto  \cite{Hikihara08} to discuss the regime of $JK<0$.  
Here we apply both types of duality to discuss the $J$-$\Jperp$-$K_{\rm ch}$ ladder \eqref{eq:H_XXX_CCI} with an {\it arbitrary} sign of $K_{\rm ch}$. 

The transformations $\Ucal$ and $\Ucals$ in Eq.\ \eqref{eq:U_Us} map the N\'eel operator \eqref{eq:O_Neel} 
to the staggered and uniform vector chiral (UVC) operators 
\begin{subequations}\label{eq:O_SVC_UVC}
\begin{align}
 \mathcal{O}_{\text{SVC}}(j) &= \frac12 \qty(\Sv_{1,j}\times\Sv_{2,j}-\Sv_{1,j+1}\times\Sv_{2,j+1})^z, \label{eq:O_SVC} \\
 \mathcal{O}_{\text{UVC}}(j) &= \frac12 \qty(\Sv_{1,j}\times\Sv_{2,j}+\Sv_{1,j+1}\times\Sv_{2,j+1})^z
\end{align}
\end{subequations}
through the relations
\begin{subequations}\label{eq:Neel_SVC_UVC}
\begin{align}
  \Ucal \mathcal{O}_{\text{N\'eel}}(j) \Ucal^\dagger &= -\mathcal{O}_{\text{SVC}}(j),\\
  (-1)^j \Ucals \mathcal{O}_{\text{N\'eel}}(j) \Ucals^\dagger &= -\mathcal{O}_{\text{UVC}}(j) .
\end{align}
\end{subequations}
These are obtained from the relations
\begin{equation}
 e^{\pm i\frac{\pi}{2} n_j^{\rm B}} \qty(S_{1,j}^\alpha-S_{2,j}^\alpha) e^{\mp i\frac{\pi}{2} n_j^{\rm B}} 
 =\mp 2 \qty(\Sv_{1,j}\times\Sv_{2,j})^\alpha. 
\end{equation}
Furthermore, $\Ucal$ maps the SD order parameter \eqref{eq:O_SD} to the SC order parameter 
\begin{equation}\label{eq:O_SC}
\begin{split}
\expval{\mathcal{O}_{\text{SC}}(j)} &= \frac14 
\langle \qty(\Sv_{1,j-1}+\Sv_{2,j-1})\cdot\qty(\Sv_{1,j}\times\Sv_{2,j}) \\
&~~~~+\qty(\Sv_{1,j-1}\times\Sv_{2,j-1})\cdot\qty(\Sv_{1,j}+\Sv_{2,j}) \\
&~~~~-\qty(\Sv_{1,j}+\Sv_{2,j})\cdot\qty(\Sv_{1,j+1}\times\Sv_{2,j+1}) \\
&~~~~-\qty(\Sv_{1,j}\times\Sv_{2,j})\cdot\qty(\Sv_{1,j+1}+\Sv_{2,j+1})\rangle 
\end{split}
\end{equation}
through the relation
\begin{align}\label{eq:SD_SC}
  \Ucal \mathcal{O}_{\text{SD}}(j) \Ucal^\dagger  = -\mathcal{O}_{\text{SC}}(j).
\end{align}
Therefore, $\Ucal$ and $\Ucals$ transform the N\'eel correlation 
into the SVC and UVC correlations, respectively, 
and $\Ucal$ transforms the SD order (correlation) into the SC order (correlation). 
Meanwhile, the plaquette and SN operators in Eqs.\ \eqref{eq:O_plaq} and \eqref{eq:O_SN} remain unchanged under $\Ucal$ or $\Ucals$. 

%############################
\begin{figure}
\includegraphics[width=0.4\textwidth]{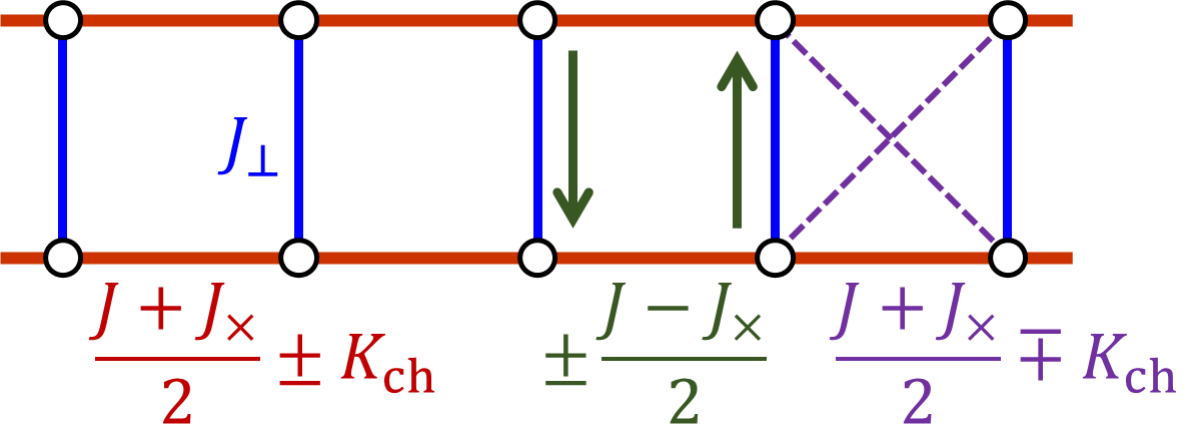}
\caption{\label{fig:laddermodel_dual}
(Color online) 
Models that result from the duality transformations \eqref{eq:U_Us} 
of the $J$-$\Jperp$-$\Jdiag$-$K_{\rm ch}$ ladder \eqref{eq:H_XXX_chex} in Fig.\ \ref{fig:laddermodel_chpm}(a). 
The upper and lower signs correspond to the cases of $\Ucal$ and $\Ucals$, respectively; 
see Eqs.\ \eqref{eq:H_chex_H12ch} and \eqref{eq:H_chpmex_duality} for the transformation rules. 
}
\end{figure}
%############################

% [ Duality transformation of the Hamiltonian ]--------------------------------------
To apply the duality, it is convenient to rewrite the Hamiltonian \eqref{eq:H_XXX_chex} 
of the $J$-$\Jperp$-$\Jdiag$-$K_{\rm ch}$ ladder as \cite{Lecheminant06,Totsuka12, Fontaine24XXZ}
\begin{equation}\label{eq:H_chex_H12ch}
\begin{split}
 H_{\rm ex}&=\frac{J+\Jdiag}{2} H_1+\frac{J-\Jdiag}{2}H_2+K_{\rm ch} H_{\rm ch}\\
 &~~~+\Jperp \sum_j \Sv_{1,j}\cdot\Sv_{2,j},
\end{split}
\end{equation}
where 
\begin{subequations}
\begin{align}
 H_1&=\sum_j \qty( \Sv_{1,j}+\Sv_{2,j} )\cdot\qty( \Sv_{1,j+1}+\Sv_{2,j+1} ),\\
 H_2&=\sum_j \qty( \Sv_{1,j}-\Sv_{2,j} )\cdot\qty( \Sv_{1,j+1}-\Sv_{2,j+1} ). 
\end{align}
\end{subequations}
The duality transformation of the Hamiltonian can then be described as the simple exchange of the coupling constants,
\begin{equation}\label{eq:H_chpmex_duality}
 \frac{J-\Jdiag}{2} \leftrightarrow \pm K_{\rm ch}, 
\end{equation}
where the upper and lower signs correspond to $\Ucal$ and $\Ucals$, respectively. 
Here, the other coupling constants $(J+\Jdiag)/2$ and $\Jperp$ in Eq.\ \eqref{eq:H_chex_H12ch} remain unchanged. 
The extended model \eqref{eq:H_XXX_chex} is thus mapped to the models in Fig.\ \ref{fig:laddermodel_dual}. 

%Under the duality transformation \eqref{eq:U_spin_chirality}, 
%$H_2$ and $H_{\rm ch}$ are interchanged while the other terms in Eq.\ \eqref{eq:H_chex_H12ch} remain invariant. 

%############################
%############################

%************************************************
\subsection{Full phase structure}\label{sec:phases}
%************************************************

% [ Dual phase diagram in the isotropic case ]--------------------------------------
The duality transformations $\Ucal$ and $\Ucals$ in Eq.\ \eqref{eq:U_Us} 
transform the upper and lower halves of Fig.\ \ref{fig:phases_chpmex}(a) into Figs.\ \ref{fig:phases_chpmex}(b) and (c), respectively, via Eq.\ \eqref{eq:H_chpmex_duality}.
Here, the N\'eel correlation is transformed into the SVC and UVC correlations in Figs.\ \ref{fig:phases_chpmex}(b) and (c), respectively, 
while the SD order (correlation) is transformed into the SC order (correlation)  in Fig.\ \ref{fig:phases_chpmex}(b). 
As seen in Fig.\ \ref{fig:laddermodel_dual}, 
the condition of weak inter-chain couplings in Eq.\ \eqref{eq:H_chpmex_weak} is now replaced by the following conditions: 
\begin{equation}\label{eq:condition_dual_chpmex}
 |\Jperp|, \bigg|\frac{J+\Jdiag}{2} \mp K_{\rm ch}\bigg|, |J-\Jdiag| \ll \frac{J+\Jdiag}{2}\pm K_{\rm ch}.
\end{equation} 
Throughout this section, the upper and lower signs correspond to the cases of Figs.\ \ref{fig:phases_chpmex}(b) and (c), respectively. 
In these figures, the $c=3/2$ RS-SC and Haldane-plaquette transitions (purple lines) occur at 
\begin{equation}\label{eq:line_WZW_dual}
 \pm \Kch = \qty(\frac12\pm \frac{1}{4\gamma}) J - \frac{\Jperp}{2} + \qty(\frac12\mp \frac{1}{4\gamma}) \Jdiag,
\end{equation}
while the $c=1/2$ Haldane-SC and plaquette-RS transitions (blue lines) occur at 
\begin{equation}\label{eq:line_Ising_dual}
 \pm \Kch = \qty(\frac12\mp \frac{1}{12\gamma}) J - \frac{\Jperp}{2} + \qty(\frac12\pm \frac{1}{12\gamma}) \Jdiag.
\end{equation}
Furthermore, changes in dominant correlation (red dotted lines) occur at 
\begin{equation}\label{eq:line_crossover_dual}
\pm \Kch = \qty(\frac12\mp \frac{1}{4\gamma}) J - \frac{\Jperp}{2} + \qty(\frac12\pm \frac{1}{4\gamma}) \Jdiag.
\end{equation}
These transition and crossover lines are obtained from Eqs.\ \eqref{eq:line_WZW}, \eqref{eq:line_Ising}, and \eqref{eq:line_crossover} via Eq.\ \eqref{eq:H_chpmex_duality}. 

% [ Combined phase diagram for the isotropic case ]--------------------------------------
By combining Figs.\ \ref{fig:phases_chpmex}(a,b,c) and setting $\Jdiag=0$, we obtain the phase diagram in Fig.\ \ref{fig:phases_chpm}. 
Here, we assumed that the phase structures in Fig.\ \ref{fig:phases_chpmex} continue qualitatively 
beyond perturbative regimes 
even when the conditions \eqref{eq:H_chpmex_weak} and \eqref{eq:condition_dual_chpmex} are nearly saturated. 
This assumption is justified {\it a posteriori} through an overall agreement 
with the numerically obtained phase diagram in Fig.\ \ref{fig:pd_iDMRG_Exact} below. 

%%%%%%%%%%%%%%%%%%%%%%%%%%%%%%%%%%%%%%%%%%%%%%%%
\section{Numerical Results}\label{sec:numer}
%%%%%%%%%%%%%%%%%%%%%%%%%%%%%%%%%%%%%%%%%%%%%%%%

% [ Section intro ]--------------------------------------
In this section, we numerically investigate the ground-state phase diagram of the $J$-$\Jperp$-$\Kch$ ladder \eqref{eq:H_XXX_CCI} 
in the $\Jperp$-$K_{\rm ch}$ plane with $J$ set to $1$. 
The obtained phase diagram (Fig.\ \ref{fig:pd_iDMRG_Exact}) is in overall agreement with the schematic phase diagram (Fig.\ \ref{fig:phases_chpm}) 
predicted by the Abelian bosonization and the duality. 
Some differences are found in the location of the plaquette-RS transition line (the blue line for $\Kch<0$): 
in the numerical result in Fig.\ \ref{fig:pd_iDMRG_Exact}, this line is always located on the $\Jperp<0$ side 
and is not bent around the $\Ucals$-self-dual line as in Fig.\ \ref{fig:phases_chpm}. 
We attribute these differences to the neglection of the marginal terms 
and a difficulty beyond a perturbative regime in field theory, 
as noted in Secs.\ \ref{sec:bos_phases} and \ref{sec:phases}. 

% [ Choice of methods ]--------------------------------------
In Fig.\ \ref{fig:pd_iDMRG_Exact}, the Ising transition lines and the crossover lines were obtained by iDMRG calculations. 
For the $c=3/2$ transitions, however, we encountered a convergence issue in iDMRG, 
which has also been noted in Refs.\ \cite{Fontaine24XXZ,Ogino21nvbs}. 
Specifically, in iDMRG with a finite bond dimension $\chi$, the $c=3/2$ transition was found to split into two transitions 
between which a spontaneous magnetic or VC order appears. 
This intermediate ordered region should be considered as an artifact due to finite $\chi$ 
as spontaneous breaking of a continuous symmetry is generally prohibited in the ground state of one-dimensional (1D) quantum lattice systems 
(unless a certain uniform susceptibility diverges) \cite{Momoi96}. 
With an increase in $\chi$, the range of this intermediate region 
decreases but does not vanish, 
obscuring the location of the transition point. 
We therefore investigate the $c=3/2$ transitions 
by exact diagonalization calculation of the polarization amplitude, as described in the following. 

%************************************************
\subsection{$c=3/2$ transitions}\label{sec:C32}
%************************************************

% [ Polarization amplitude ]--------------------------------------
The polarization amplitude $z_{\rm rung}$ 
\cite{Resta98, RestaSorella99, AligiaOrtiz99, AligiaHallberg00, NakamuraTodo02, NakamuraVoit02, Nakamura03, FuruyaSato21, Tasaki18}
corresponds to the expectation value of the Lieb-Schultz-Mattis twist operator \cite{Lieb61,Oshikawa97}. 
For a two-leg ladder of length $L$ with the periodic boundary condition, it is defined as \cite{Laeuchli03,Hikihara10,Nakamura03, FuruyaSato21, Ogino22}
\begin{equation}\label{eq:z_rung}
 z_{\rm rung} = \Bigg\langle \exp\qty( i\frac{2\pi}{L} \sum_{n=1}^2 \sum_{j=1}^L j S_{n,j}^z) \Bigg\rangle .
\end{equation}
In the Abelian bosonization, the polarization amplitude can be expressed as \cite{NakamuraVoit02, Nakamura03, FuruyaSato21} 
\begin{equation}\label{eq:z_rung_phi}
 z_{\rm rung}\propto -\Big\langle \cos\qty(2\sqrt{2\pi} \phi_+) \Big\rangle . 
\end{equation}
The sign of this quantity therefore reflects the locking position of $2\sqrt{2\pi} \phi_+$, 
and a sign change is expected to occur across the $c=3/2$ transitions in Fig.\ \ref{fig:phases_chpmex}(a) and their dual counterparts. 

% [ Results on the polarization amplitude ]--------------------------------------
In our previous paper \cite{Fontaine24XXZ}, the polarization amplitude has been calculated for the $\Jperp,K_{\rm ch}>0$ part of the phase diagram. 
Here, we extend this analysis for $\Jperp, \Kch<0$. 
Exact diagonalization was performed with the Python package QuSpin \cite{QuSpin17, QuSpin19} for ladder lengths $L = 8$, $10$, $12$, and $14$. 
Finite-size estimates of the $c=3/2$ transitions are obtained from the zeros of $z_{\rm rung}$ (purple triangles in Fig.\ \ref{fig:pd_iDMRG_Exact}),  
and they can be extrapolated to the infinite-size limit using a linear function of $L^{-2}$. 
This fitting works well for $\Jperp\le 1$, and the extrapolated transition points are interpolated by purple solid lines in Fig.\ \ref{fig:pd_iDMRG_Exact}. 

%############################
\begin{figure}
\includegraphics[width=0.4\textwidth]{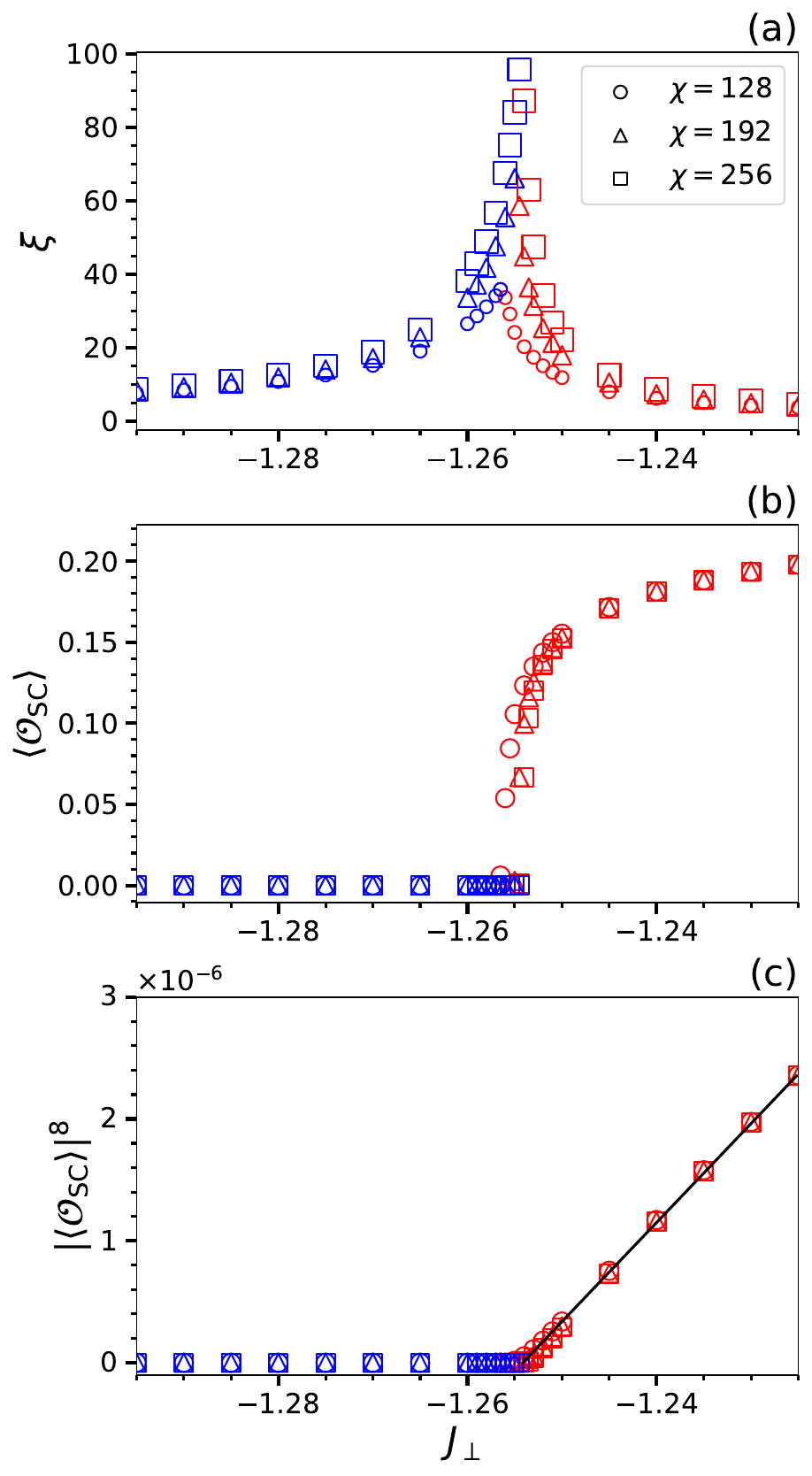}
\caption{\label{fig:H_SC_cor_odp}
(Color online) 
(a) Correlation length $\xi$, 
(b) the SC order parameter $\langle\mathcal{O}_{\rm{SC}}\rangle$ in Eq.\ \eqref{eq:O_SC}, and 
(c) $\langle \mathcal{O}_{\rm{SC}}\rangle^8$, calculated by iDMRG across the Haldane-SC transtion at $K_{\rm ch} = 0.8$. 
Blue (red) symbols indicate data calculated using an initial point deep in the Haldane (SC) phase. 
Three different symbol shapes correspond to different bond dimensions $\chi$, as indicated in the legends. 
In (a), a maximum of $\xi$ occurs at $J_{\perp,c}\approx -1.2565, -1.2550$, and $-1.2545$ (with a precision of 0.0005) 
for $\chi = 128$, $192$, and $256$, respectively. 
In (c), the solid black line is the linear fitting done in the range not too close to the transition  
where the $\chi$-dependence of $\langle\mathcal{O}_{\rm{SC}}\rangle$ is almost inperceptible. 
The intercept with the horizontal axis gives $J_{\perp,c}\approx -1.2541$.  
}
\end{figure}
%############################

%************************************************
\subsection{Ising transitions}\label{sec:iDMRG_ising}
%************************************************

% [ iDMRG ]--------------------------------------
The Ising transitions were investigated by iDMRG using the C++ library ITensor \cite{itensor,itensor-r0.3}. 
In the iDMRG algorithm \cite{McCulloch07,McCulloch08}, one represents a many-body wave function in the infinite 1D system 
by a periodic matrix product state (MPS) with a bond dimension $\chi$. 
For all iDMRG calculations done in this paper, we used a MPS with a four-site (i.e., two-rung) period 
and three different bond dimensions ($\chi = 128,192, 256$). 

% [ Convergence issues on the Ising transitions ]--------------------------------------
Near the transitions, we encountered convergence issues such as a dependence on the initial state, 
which tended to enlarge symmetric phases. 
To remedy this problem, we used a point deep in the symmetric (symmetry broken) phase as the initial state of the optimization process, 
and we represent the corresponding results in blue (red) in Figs.\ \ref{fig:H_SC_cor_odp} and \ref{fig:Pl_RS_cor_odp} shown below. 
For each value of $J_\perp$, $K_{\rm ch}$, and $\chi$, we only retained the point with the lowest energy. 
 
% [ Results on Ising phase transition points ]--------------------------------------
In the phase diagram in Fig.\ \ref{fig:pd_iDMRG_Exact}, Ising transition points (blue circles) are determined by iDMRG from the (local) maximum of the correlation length. 
Here, the correlation length is obtained from the eigenvalue of the transfer matrix, as described in Ref.\ \cite{Fontaine24XXZ}.  
The transition points estimated with different $\chi$ are shown by circles of different sizes, and they are almost overlapping. 
Therefore, the data points for the largest bond dimension ($\chi=256$) are expected to give reasonably accurate phase boundaries, 
and they are interpolated by blue solid lines as a guide for the eye in the figure. 

% [  ]--------------------------------------
Below we focus on the Haldane-SC and plaquette-RS transitions as representative cases. 
The Haldane-SD and plaquette-RS transitions have been analyzed in related ladder models with an inter-leg DDI \cite{Hijii13,ChenJH19,Robinson19}. 
Different behaviors of the plaquette-RS transition lines in the present model and the models of Ref.\ \cite{Hijii13,ChenJH19,Robinson19} 
have been discussed from a field-theoretical point of view in Sec.\ \ref{sec:bos_phases}. 

%############################
\begin{figure}
\includegraphics[width=0.45\textwidth]{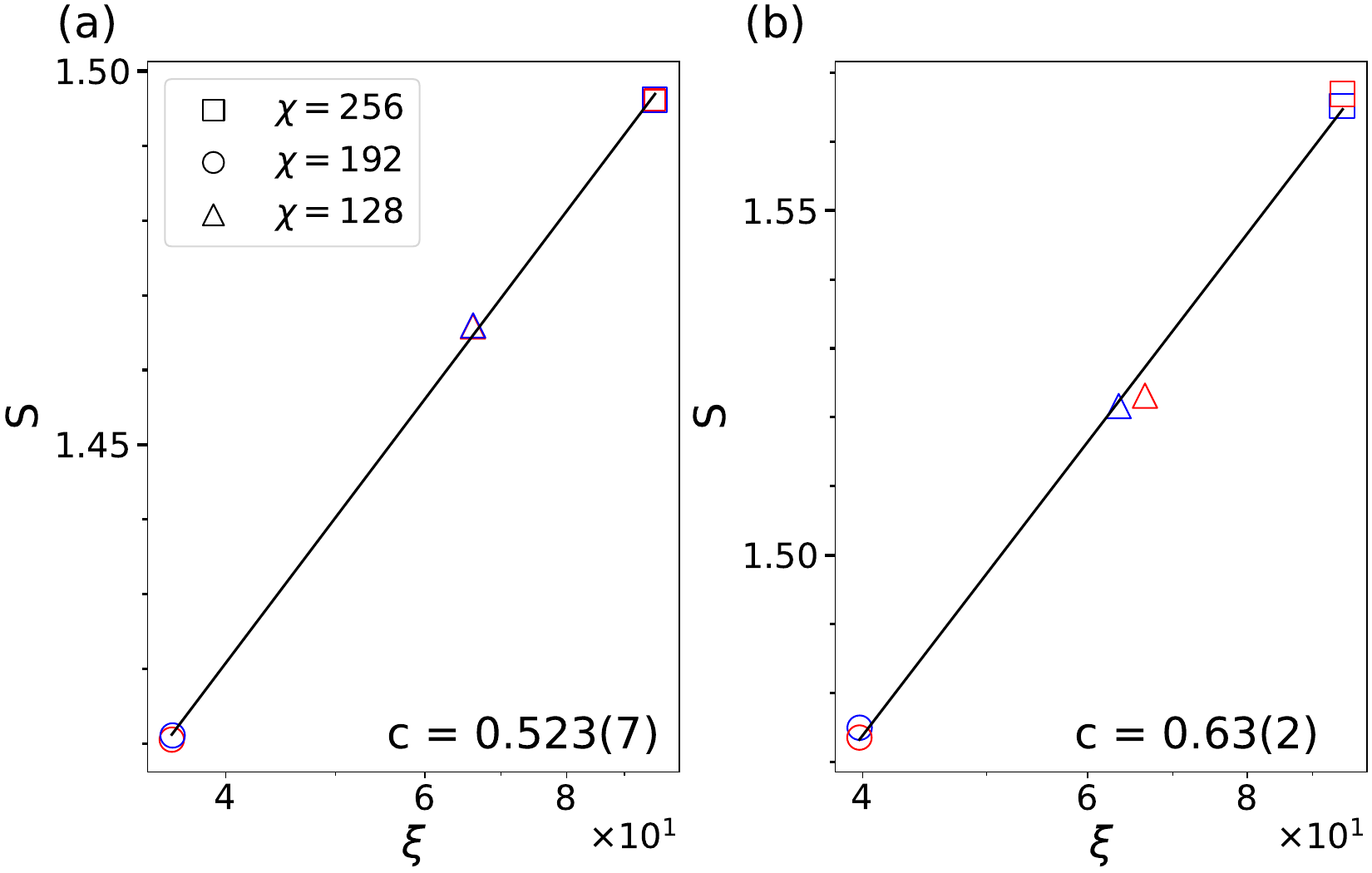}
\caption{\label{fig:CC_Ising}
(Color online) 
Half-chain EE $S$ versus the correlation length $\xi$, calculated by iDMRG at (a) the Haldane-SC transition at $K_{\rm ch} = 0.8$ in Fig.\ \ref{fig:H_SC_cor_odp} 
and (b) the plaquette-RS transition at $K_{\rm ch} =- 0.8$ in Fig.\ \ref{fig:Pl_RS_cor_odp}. 
For each bond dimension $\chi$, the point with the highest correlation length in each phase is plotted. 
The same symbols and colors as in Figs.\ \ref{fig:H_SC_cor_odp} and \ref{fig:Pl_RS_cor_odp} are used. 
A logarithmic scale is used for the horizontal axis. 
Solid lines show the fitting with Eq.\ \eqref{eq:CC}, from which the central charge $c$ is estimated as shown in each panel. 
Here, the number in parentheses in estimated $c$ shows an error in the last digit that is associated with the linear fitting.  
The results are in reasonable agreement with the Ising universality class with $c=1/2$. 
}
\end{figure}
%############################

%############################
\begin{figure}
\includegraphics[width=0.4\textwidth]{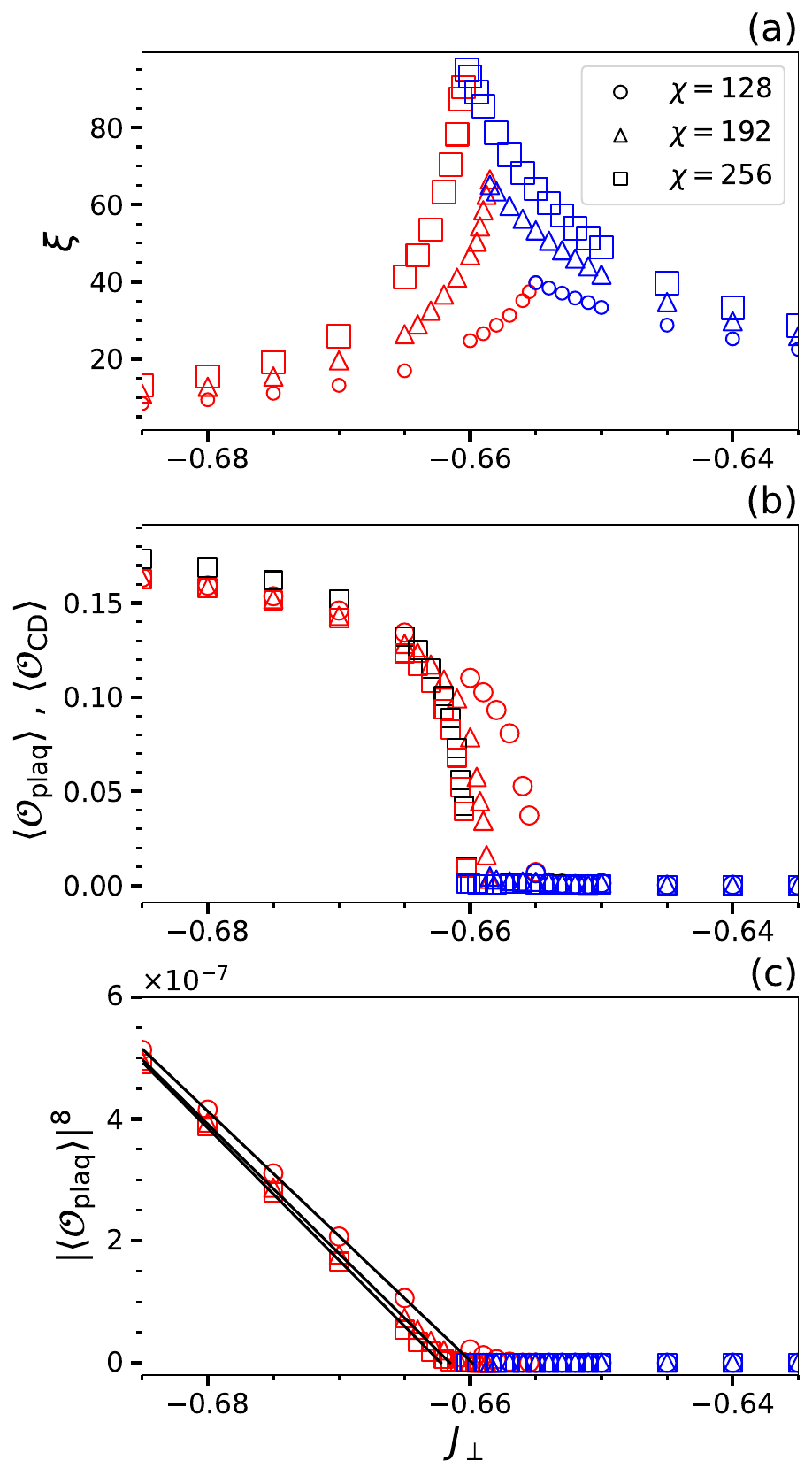}
\caption{\label{fig:Pl_RS_cor_odp}
(Color online) 
(a) Correlation length $\xi$, 
(b) the plaquette order parameters $\langle\mathcal{O}_{\rm{plaq}}\rangle$ in Eq.\ \eqref{eq:O_plaq} (red and blue)
and the CD order parameter $\langle\mathcal{O}_{\rm{CD}}\rangle$ in Eq.\ \eqref{eq:O_CD} (black), 
(c) $\langle \mathcal{O}_{\rm{plaq}}\rangle^8$, calculated by iDMRG across the plaquette-RS transtion at $K_{\rm ch} = -0.8$. 
Red (blue) symbols indicate data points calculated using an initial point in the plaquette (RS) phase.  
In (a), a maximum of $\xi$ occurs at $J_{\perp,c}\approx -0.65500$, $-0.65850$, and $-0.66025$ 
(with a precision of 0.00025) 
for $\chi = 128$, $192$, and $256$, respectively. 
In (b), the CD order parameter is plotted only for $\chi = 256$ for reference.
In (c), the solid lines show the linear fitting done not too close to the transition for each $\chi$. 
The intercept with the horizontal axis gives $J_{\perp}\approx -0.6598$, $-0.6615$, and $-0.6622$ for $\chi = 128$, $192$, and $256$, respectively.
}
\end{figure}
%############################

%############################
\begin{figure*}
\includegraphics[width=0.8\textwidth]{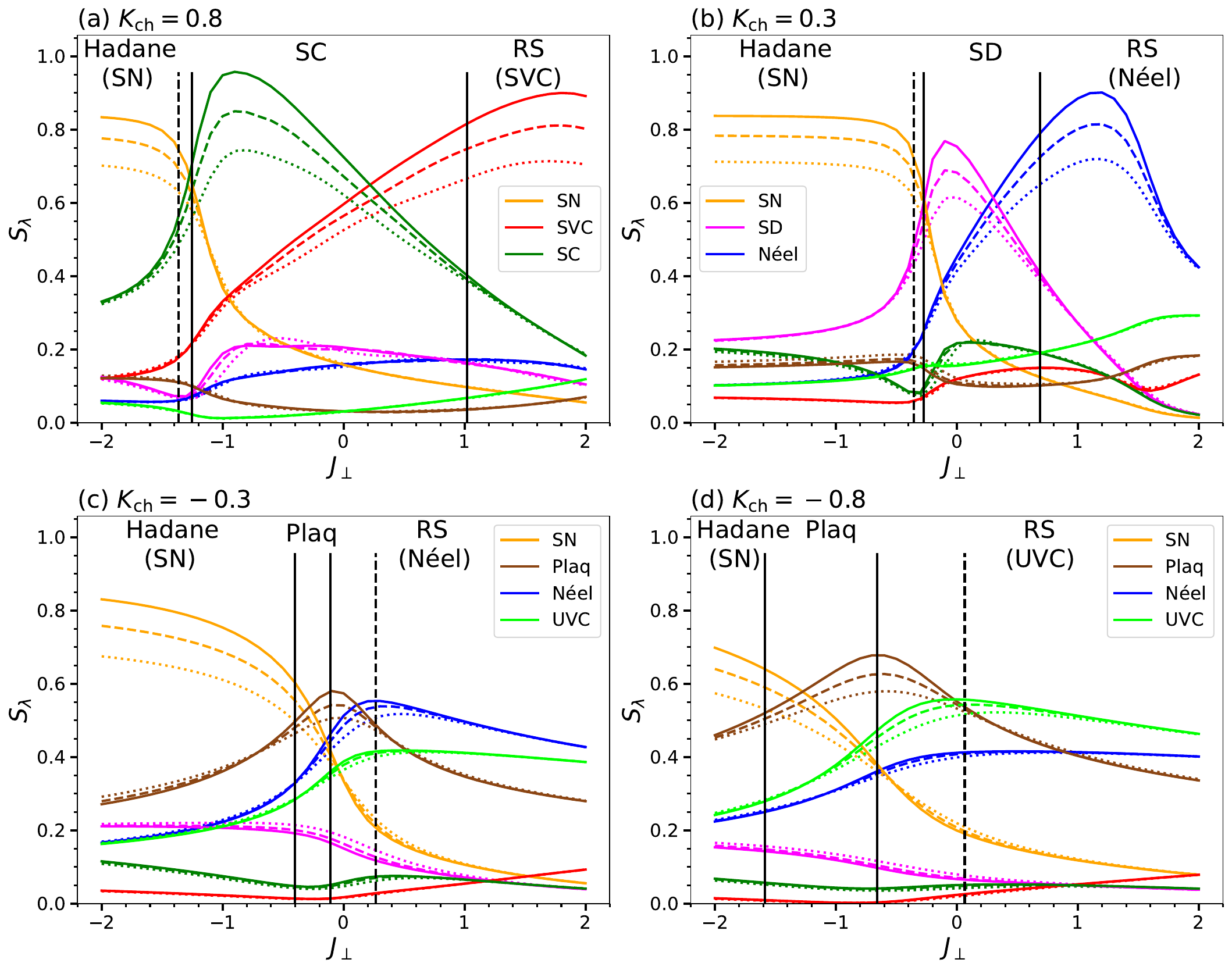}
\caption{\label{fig:DM_cor}
(Color online) 
Structure factors in Eqs.\ \eqref{eq:SSF} and \eqref{eq:SF_UVC} calculated by exact diagonalization at (a,d) $\Kch=\pm 0.8$ and (b,c) $\Kch=\pm 0.3$. 
Data for the ladder lengths $L = 8$, $10$, and $12$ are shown by dotted, dashed, and solid lines, respectively. 
%In the legends, Néel, SN, SD, SVC, SC, and Plaq refer to the corresponding staggered structure factors while UVC refers to its uniform structure factor. 
Vertical solid lines indicate the $c=3/2$ and Ising transition points determined in Secs.\ \ref{sec:C32} and \ref{sec:iDMRG_ising}. 
Vertical dashed lines indicate the crossover points determined in Sec.\ \ref{sec:cross}.  
On the top of each figure, we indicate the names of the phases and the expected dominant correlations in parenthesis. 
Dominant correlation in the region between the crossover and transition lines is the same as the one in the symmetry-broken phase. 
}
\end{figure*}
%############################

%************************************************
\subsubsection{Haldane-SC transition}\label{sec:H-SC}
%************************************************

% [ Correlation length and order parameter ]--------------------------------------
The Haldane-SC transition is the $\Ucal$-dual counterpart to the Haldane-SD transition, and it is expected to belong to the Ising universality class. 
In Fig.\ \ref{fig:H_SC_cor_odp}, we analyze the transition at $K_{\rm ch} = 0.8$ as a representative case. 
As seen in Fig.\ \ref{fig:H_SC_cor_odp}(a), the maximum of the correlation length $\xi$ grows with the bond dimension $\chi$, 
suggesting a continuous transition. 
The maximum of $\xi$ is associated with the onset of the SC order as seen in Fig.\ \ref{fig:H_SC_cor_odp}(b). 

% [ Central charge ]--------------------------------------
To investigate the nature of the transition, we employed two methods. 
According to conformal field theory, the half-chain entanglement entropy (EE) $S$ and the correlation length $\xi$ 
fulfill the following relationship \cite{Calabrese04,Pollmann09}:
\begin{equation}\label{eq:CC}
 S = \frac{c}{6} \ln \xi + S_0,
\end{equation}
where $c$ is the central charge and $S_0$ is a non-universal constant. 
In Fig.\ \ref{fig:CC_Ising}(a), we plot the relation between $S$ and $\xi$ calculated by iDMRG at the Haldane-SC transition. 
The data well obey the scaling form \eqref{eq:CC}, and the central charge $c=0.523(7)$ 
obtained from the fitting is in good agreement with the Ising universality class with $c=1/2$. 

% [ Order parameter to the 8th power ]--------------------------------------
For the Ising universality class, the critical exponent for the order parameter is known to be $\beta=1/8$. 
The order parameter to the eighth power, $\langle\mathcal{O}_{\rm{SC}}\rangle^8$, is thus expected to behave linearly around the transition. 
As seen in Fig.\ \ref{fig:H_SC_cor_odp}(c), $\langle\mathcal{O}_{\rm{SC}}\rangle^8$ is indeed well fitted by a linear function 
in the regime where the $\chi$-dependence of $\langle\mathcal{O}_{\rm{SC}}\rangle$ is almost imperceptible. 
%The intercept with the horizontal axis gives an accurate estimate  $J_{\perp,c}\approx -1.2541$ of the transition point. 

%************************************************
\subsubsection{Plaquette-RS transition}\label{sec:PL-RS}
%************************************************

% [ Correlation length order parameters ]--------------------------------------
In Fig.\ \ref{fig:Pl_RS_cor_odp}, we analyze the plaquette-RS transition at $K_{\rm ch}  = -0.8$ as a representative case. 
The correlation length $\xi$ in Fig.\ \ref{fig:Pl_RS_cor_odp} (a) shows a maximum, 
which is associated with the onset of the plaquette order (red symbols) in Fig.\ \ref{fig:Pl_RS_cor_odp}(b). 
In Fig.\ \ref{fig:Pl_RS_cor_odp}(b), the CD order parameter \eqref{eq:O_CD} for $\chi=256$ is also plotted in black for reference. 
The plaquette and CD order parameters show a very similar behavior, and they indeed start to grow at the same $\Jperp$ in Fig.\ \ref{fig:Pl_RS_cor_odp}(b). 
Below we shall employ the plaquette order parameter \eqref{eq:O_plaq}, which is invariant under $\Ucals$, for the characterization of the phase. 
In Fig.\ \ref{fig:Pl_RS_cor_odp}(c), the linear fitting of $\langle \mathcal{O}_{\rm{plaq}}\rangle^8$ is performed for each $\chi$. 
While the $\chi$-dependence is not negligible, the linear fitting works well especially for $\chi=192$ and $256$, 
and the lines for these $\chi$'s are very close, giving a support for the Ising universality class. 
The transition points estimated from the intercept are also given in the caption of Fig.\ \ref{fig:Pl_RS_cor_odp} for reference. 
As the linear fitting in Fig.\ \ref{fig:Pl_RS_cor_odp}(c) could be affected by the choice of the fitting region, 
we employed the simpler estimates from the maximum of $\xi$ for the largest $\chi$ in the phase diagram of Fig.\ \ref{fig:pd_iDMRG_Exact}. 

% [  ]--------------------------------------
%In Fig.\ \ref{fig:H_SC_cor_odp} and \ \ref{fig:Pl_RS_cor_odp}(c), the intercept of the linear fitting with the horizontal axis gives an alternative estimate of the transition point. However, arbitrariness in the choice of the fitting region deeply affects the estimated value of the transition point. Therefore, we employed the simpler estimates from the maximum of $\xi$ for the largest $\chi$ in the phase diagram in Fig.\ \ref{fig:pd_iDMRG_Exact}. 

% [ Central chargee ]--------------------------------------
In Fig.\ \ref{fig:CC_Ising}(b), we plot the relation between $S$ and $\xi$ at the maxima in Fig.\ \ref{fig:Pl_RS_cor_odp}(a). 
The central charge $c=0.63(2)$ calculated from the $S$-$\xi$ relation agrees reasonably with the Ising universality class with $c=1/2$.  

%************************************************
\subsection{Dominant correlation}\label{sec:DC}
%************************************************

While the RS and Haldane phases are featureless and do not spontaneously break any symmetry, 
they can be divided into several regions with different dominant correlations, 
as predicted in the field-theoretical analysis in Sec.\ \ref{sec:EFT_XXX}. 
Here we analyze structure factors and correlation functions to discuss changes in dominant correlation. 

%In previous subsections, the phase transitions and their nature were numerically investigated. In this session, we will focus on the nature of each phase. In a 1D quantum system, strong quantum fluctuation prevents the existence of phases that break continuous symmetry. Long-range order is prevented but phases can still be characterized by their dominant correlation. 

%************************************************
\subsubsection{Structure factors}\label{sec:SF}
%************************************************

% [ Structure factors ]--------------------------------------
As a guide to investigate dominant correlation,  
we first examine the staggered structure factors 
\begin{equation}\label{eq:SSF}
 S_\lambda = \frac1L \left\langle \bigg[ \sum_j (-1)^j \mathcal{O}_\lambda (j) \bigg]^2 \right\rangle,
\end{equation}
for $\lambda=\text{N\'eel}$, $\mathrm{SN}$, $\mathrm{SD}$, $\mathrm{plaq}$, $\text{SC}$, and $\text{SVC}$, 
%$\lambda=\mathrm{SD}$, $\text{SC}$, $\mathrm{plaq}$, $\mathrm{SN}$, $\text{N\'eel}$, and $\text{SVC}$, 
as well as the UVC structure factor
\begin{equation}\label{eq:SF_UVC}
 S_\text{UVC} = \frac1L \left\langle \bigg[ \sum_j \mathcal{O}_\text{UVC}(j) \bigg]^2 \right\rangle .
\end{equation}
The structure factors calculated by exact diagonalization 
are plotted as a function of $J_\perp$ for four representative values of $K_{\rm ch}$ in Fig.\ \ref{fig:DM_cor}.  
In general, asymptotically for large $L$, the structure factor increases linearly with $L$ for a long-range order, 
and converges to a finite value for a short-range order. 
Here we do not try to distinguish between long- and short-range orders, 
but aim to roughly identify the development of different correlations in the phase diagram. 

% [ Kch = -0.3, -0.8 ]--------------------------------------
For $K_{\rm ch} = -0.3$ in Fig.\ \ref{fig:DM_cor}(c), we find that the SN, plaquette, and N\'eel structure factors become successively the largest with increasing $\Jperp$. 
In particular, a large value of the plaquette structure factor continues beyond the plaquette-RS transition point $\Jperp=-0.110(5)$, 
and thus the change in dominant correlation from plaquette to N\'eel seems to occur {\it inside} the RS phase, 
in consistency with the field-theoretical prediction in Fig.\ \ref{fig:phases_chpm}. 
The result for $K_{\rm ch} = -0.8$ in Fig.\ \ref{fig:DM_cor}(d) can roughly be viewed as the $\Ucals$-dual counterpart to Fig.\ \ref{fig:DM_cor}(c). 
Instead of the N\'eel one, the UVC structure factor becomes the largest in the rightmost region of Fig.\ \ref{fig:DM_cor}(d). 
We note that the N\'eel and UVC structure factors should coincide on the $\Ucals$-self-dual line at $\Kch=-1/2$. 

% [ Kch = 0.3, 0.8 ]--------------------------------------
For $K_{\rm ch}=0.3$ in Fig.\ \ref{fig:DM_cor}(b), 
the SN, SD, and N\'eel structure factors become successively the largest with increasng $\Jperp$. 
Here, the change in dominant correlation from SN to SD seems to occur near the Haldane-SD transition point $\Jperp=-0.660(5)$, 
and its occurrence {\it inside} the Haldane phase as predicted in Fig.\ \ref{fig:phases_chpm} is not clear, 
requiring a more detailed analysis in Sec.\ \ref{sec:cross}. 
The result for $K_{\rm ch} = 0.8$ in Fig.\ \ref{fig:DM_cor}(a) can roughly be viewed as the $\Ucal$-dual counterpart to Fig.\ \ref{fig:DM_cor}(b). 
Instead of the SD and N\'eel correlations in Fig.\ \ref{fig:DM_cor}(b), the SC and SVC correlations develop in Fig.\ \ref{fig:DM_cor}(a). 
The SD and SC or N\'eel and SVC structure factors should coincide on the $\Ucal$-self-dual line at $\Kch=1/2$. 
We note that  in the exact RS state for $\Kch=1/2$ and $\Jperp>2$ (see Fig.\ \ref{fig:phases_chpm}), 
all local correlation functions [as the ones in Eqs.\ \eqref{eq:corrfn_lambda} and \eqref{eq:corrfn_UVC} below] 
strictly disappear unless the two operators overlap.

%############################
 \begin{figure}
\includegraphics[width=0.5\textwidth]{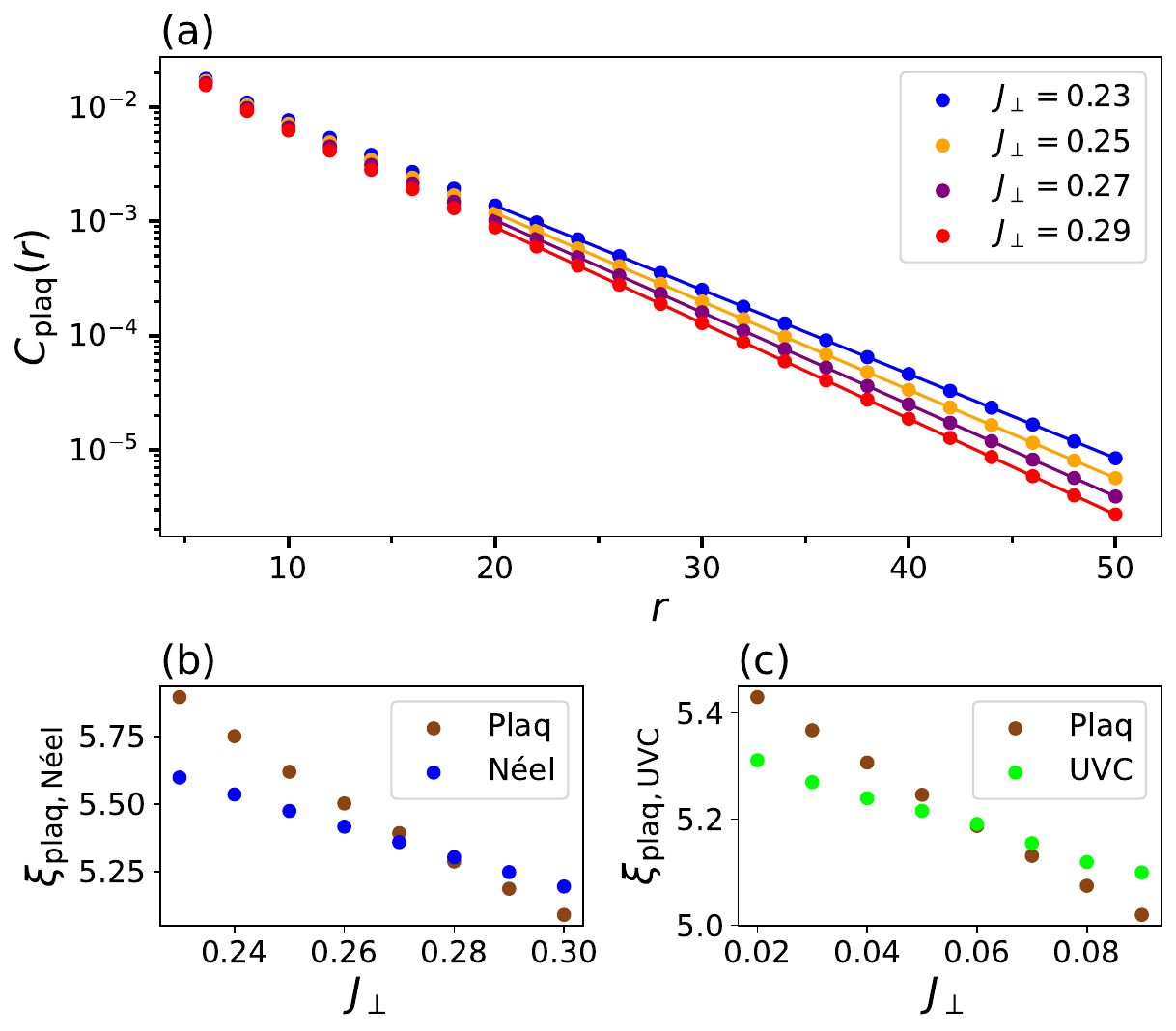}
\caption{\label{fig:cross_1}
(Color online) 
(a) Plaquette correlation function $C_{\rm plaq}(r)$ in Eqs.\ \eqref{eq:corrfn_lambda} (in logarithmic scales) calculated by iDMRG with $\chi = 128$ 
for fixed $K_{\rm ch} = -0.3$ and for representative values of $J_\perp$ shown in the legend. 
Lines show the fitting with the exponential form $A\exp(-r/\xi_{\rm plaq})$ for $20\le r\le 50$, 
from which the correlation length $\xi_{\rm plaq}$ can be extracted. 
(b) Correlation lengths extracted from the plaquette and N\'eel correlation functions at $\Kch= -0.3$. 
(c) Correlation lengths from the plaquette and UVC correlation functions at $\Kch= -0.8$. 
The crossover points inside the RS phase are estimated to be $\Jperp\approx 0.28$ and $0.06$ in (b) and (c), respectively. 
}
\end{figure}
%############################

%************************************************
\subsubsection{Correlation functions}\label{sec:cross}
%************************************************

%The field-theoretical analysis of Sec.\ \ref{sec:EFT_XXX} predicts the Haldane and Rung Singlet phases can be divided between different regions according to the dominant correlation. The analysis of the structure factor determined by exact diagonalization can help to determine dominant correlations however strong finite size effect makes the precise localization of the crossover from one region to the other difficult. To overcome this difficulty we perform a direct comparison of the correlation function calculated using iDMRG. At the crossover, the correlation function remains finite and the ground state can be closely approximated with a small bound dimension ($\chi = 128$ in our case). 

% [ Correlation functions ]--------------------------------------
To discuss changes in dominant correlation more closely, we examine the staggered correlation functions 
\begin{equation}\label{eq:corrfn_lambda}
 C_\lambda(r)=(-1)^r \expval{\mathcal{O}_\lambda (0)\mathcal{O}_\lambda(r)}
\end{equation}
for $\lambda=\text{N\'eel}$, $\mathrm{SN}$, $\mathrm{SD}$, $\mathrm{plaq}$, $\text{SC}$, and $\text{SVC}$, 
as well as the UVC correlation function 
\begin{equation}\label{eq:corrfn_UVC}
 C_\text{UVC}(r)=\expval{\mathcal{O}_\text{UVC} (0)\mathcal{O}_\text{UVC} (r)}. 
\end{equation}

% [ Plaquette-Neel crossover ]--------------------------------------
In Fig.\ \ref{fig:cross_1}(a), we show the plaquette correlation function $C_{\rm plaq}(r)$ at $\Kch= -0.3$ inside the RS phase as a representative case. 
At large $r$, the data are fit well by the exponential function, from which the correlation length $\xi_{\rm plaq}$ is extracted. 
In Fig.\ \ref{fig:cross_1}(b), we compare the correlation lengths extracted from the plaquette and N\'eel correlation functions; 
these lengths are expected to correspond to $v/|m_{\rm s}|$ and $v/m_{\rm t}$, respectively, in Eq.\ \eqref{eq:sigma_corr_decay}. 
We find $\xi_{\rm plaq}=\xi_{\text{N\'eel}}$ at $\Jperp\approx 0.28$, wihch can be identified as the crossover point. 
Similaly, in Fig.\ \ref{fig:cross_1}(c), the plaquette-UVC crossover point at $K_{\rm ch}=-0.8$ inside the RS phase is estimated to be $\Jperp\approx 0.06$. 
In these ways, we have determined the crossover points (red squares) in Fig.\ \ref{fig:pd_iDMRG_Exact}. 

\section{Summary and Outlook}\label{sec:summary}
%%%%%%%%%%%%%%%%%%%%%%%%%%%%%%%%%%%%%%%%%%%%%%%%

In this paper, we have studied the ground-state phase diagram of a spin-$\frac12$ Heisenberg ladder with a CCI in Eq.\ \eqref{eq:H_XXX_CCI} 
by means of the Abelian bosonization, fermionization, two types of spin-chirality duality, iDMRG, and exact diagonalization. 
We have obtained a comprehensive phase diagram that consists of the RS phase, the Haldane phase, and three non-magnetic ordered phases (SD, SC, and plaquette), 
as shown in Figs.\ \ref{fig:phases_chpm} and \ref{fig:pd_iDMRG_Exact}. 
Despite their featureless nature, the RS and Haldane phases can be divided into several regions 
with different dominant correlations in terms of spin, dimer, or chirality. 
These results demonstrate a rich interplay between spin and chirality in the present system. 

An important future issue is to determine the phase diagram of 
a spin-$\frac12$ Heisenberg ladder with a four-spin ring exchange in Fig.\ \ref{fig:laddermodel_chpm}(b). 
While the cases of $\Jperp=J$ \cite{Laeuchli03, Hikihara03, Hijii02, Hijii03, Capponi13} 
and $\Jperp=-J<0$ \cite{Capponi13} have been analyzed previously, 
a rich phase structure  as found in the present work is expected in the $\Jperp$-$K$ plane. 
It will also be interesting to investigate dominant correlation in a variety of SPT and trivial phases 
in extended ladder systems \cite{Ogino21nvbs,Ogino21spt,Ogino22,Fontaine24XXZ,LiuZX12,Manmana13, Fuji15, Fuji16}, 
for which bosonization and fermionization analyses as in Sec.\ \ref{sec:EFT_XXX} are expected to provide useful guides. 

%%%%%%%%%%%%%%%%%%%%%%%%%%%%%%%%%%%%%%%%%%%%%%%%
% Acknowledgements
%%%%%%%%%%%%%%%%%%%%%%%%%%%%%%%%%%%%%%%%%%%%%%%%

\begin{acknowledgments}
The authors thank Koudai Sugimoto for stimulating discussions. 
This research was supported 
by MEXT KAKENHI Grant-in-Aid for Transformative Research Areas A ``Extreme Universe'' No.\ JP24H00970, 
by JSPS KAKENHI Grant No.\ JP18K03446, 
and by the Center of Innovations for Sustainable Quantum AI (JST Grant No.\ JPMJPF2221). 
M.F. was also supported by Keio ``Design the Future'' Award for International Students. 
\end{acknowledgments}

\appendix

%%%%%%%%%%%%%%%%%%%%%%%%%%%%%%%%%%%%%%%%%%%%%%%%
\section{Phase Diagram in the Easy-Axis Regime}\label{sec:easy-axis}
%%%%%%%%%%%%%%%%%%%%%%%%%%%%%%%%%%%%%%%%%%%%%%%%

\newcommand{\Scal}{{\cal S}}

%############################
\begin{figure}
\includegraphics[width=0.45\textwidth]{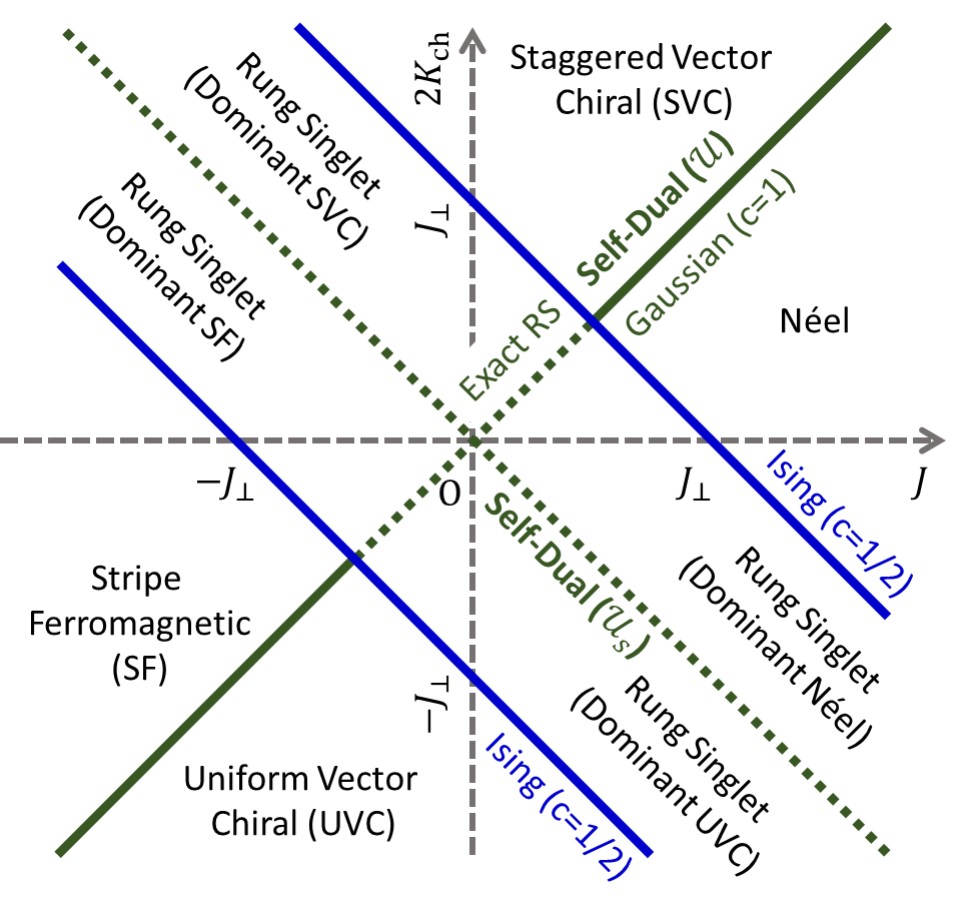}
\caption{\label{fig:phases_chpm_Jpm_Ising}
(Color online) 
Ground-state phase diagram of the XXZ-CCI ladder [defined in Eqs.\ \eqref{eq:H_XXX_CCI} and \eqref{eq:Jperp_XXZ}] with $\Jperp>0$ in the easy-axis regime, 
shown in the $J$-$2\Kch$ plane. 
This is based on the effective XY Hamiltonian \eqref{eq:H_XY}, which is exactly solvable. 
The four types of orders are shown in Fig.\ \ref{fig:ladderstates_chpm}(g,h,i,j). 
Ising transitions with the central charge $c=1/2$ occur along the blue lines between the RS and ordered phases. 
The model is self-dual under $\Ucal$ or $\Ucals$ at $K_{\rm ch}=J/2$ and $K_{\rm ch}=-J/2$, respectively (green lines). 
Along the $\Ucal$-self-dual line with $|J|<\Jperp/2$, the ground state is given by the exact RS state \cite{Fontaine24XXZ}. 
}
\end{figure}
%############################

%############################
\begin{figure}
\includegraphics[width=0.5\textwidth]{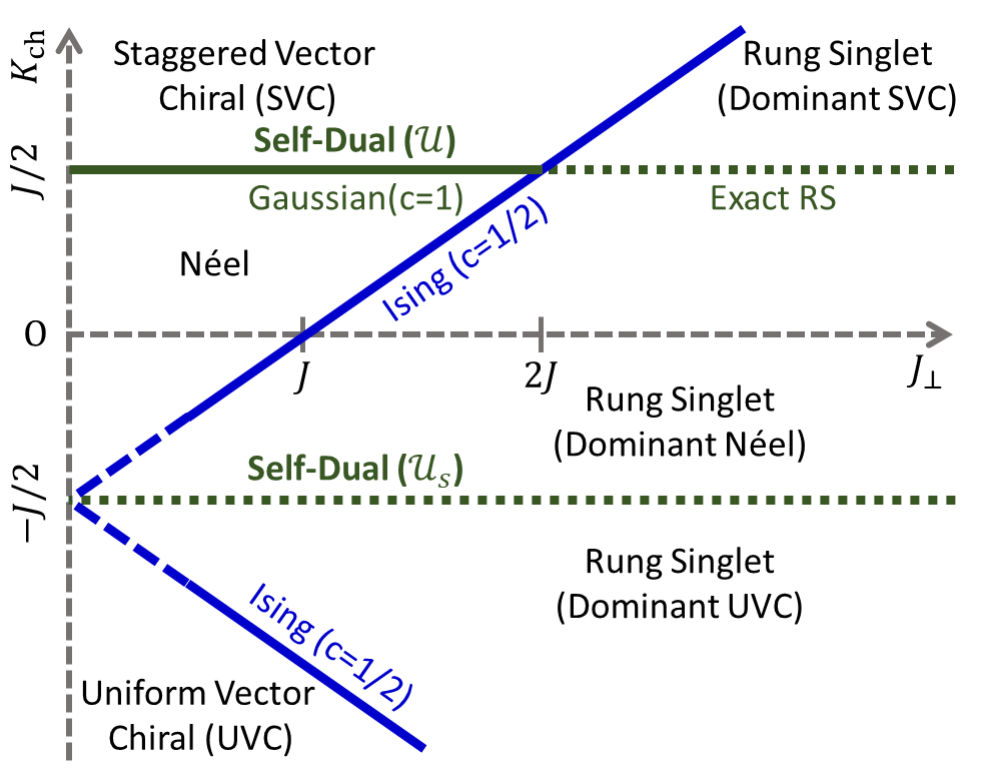}
\caption{\label{fig:phases_chpm_Ising}
(Color online) 
Ground-state phase diagram of the XXZ-CCI ladder with $J,\Jperp>0$ in the easy-axis regime, 
shown in the $\Jperp$-$\Kch$ plane. 
This is obtained by converting the right half of Fig.\ \ref{fig:phases_chpm_Jpm_Ising}. 
The case of $\Kch>0$ has been discussed previously in Ref.\ \cite{Fontaine24XXZ}. 
}
\end{figure}
%############################

Here we consider an XXZ extension of the $J$-$\Jperp$-$K_{\rm ch}$ ladder \eqref{eq:H_XXX_CCI} with $\Jperp>0$, 
and discuss the phase diagram in the easy-axis regime. 
This gives a simple picture for the emergence of spin and VC correlations in this system. 
The case of $J,\Kch>0$ has been discussed in Ref.\ \cite{Fontaine24XXZ}, and here we extend it for arbitrary signs of $J$ and $\Kch$. 

Specifically, we replace the $\Jperp$ term in Eq.\ \eqref{eq:H_XXX_CCI} by the XXZ interaction 
\begin{equation}\label{eq:Jperp_XXZ}
 \Jperp \sum_j \left( S_{1,j}^x S_{2,j}^x + S_{1,j}^y S_{2,j}^y + \Delta S_{1,j}^z S_{2,j}^z \right), 
\end{equation}
and consider the regime $\Jperp(\Delta-1)\gg |J|,|\Kch|$. 
In this regime, on each rung, it is reasonable to restrict ourselves to the subspace spanned 
the two lowest eigenstates
\begin{subequations}
\begin{align}
 |s\rangle &= \frac{1}{\sqrt{2}} (|\uparrow\downarrow\rangle-|\downarrow\uparrow\rangle),\\ 
 |t_z\rangle &= \frac{1}{\sqrt{2}} (|\uparrow\downarrow\rangle+|\downarrow\uparrow\rangle)
\end{align}
\end{subequations}
of the local $\Jperp$ interaction. 
In this subspace, we introduce the pseudospin-$\frac12$ operator 
\begin{equation}\label{eq:Scal}
 \Scal^\alpha_j  = \frac12 
 \left(  |t_z^{[j]}\rangle,  |s^{[j]}\rangle \right) \sigma^\alpha 
 \begin{pmatrix}  \langle t_z^{[j]}| \\  \langle s^{[j]}| \end{pmatrix}~~
 (\alpha=x,y,z),
\end{equation}
where $(\sigma^x,\sigma^y,\sigma^z)$ are the Pauli matrices and $[j]$ indicates the concerned rung. 
We then obtain the low-energy effective Hamiltonian in the form of the XY chain in a magnetic field \cite{Fontaine24XXZ}
\begin{equation}\label{eq:H_XY}
 H = \sum_j \qty( 2J\Scal_j^x \Scal_{j+1}^x + 4K_{\rm ch} \Scal_j^y \Scal_{j+1}^y ) + \Jperp \sum_j \Scal_j^z + \text{const.}
\end{equation}

The effective Hamiltonian \eqref{eq:H_XY} is exactly solvable, and its ground-state phase diagram is well-known \cite{Dutta15book, BarouchMcCoy71}. 
When $\Jperp>|J+2K_{\rm ch}|$, the pseudosopin model \eqref{eq:H_XY} is in a ``paramagnetic phase'', 
in which pseudospins are mainly polarized in the $-z$ direction and no spontaneous symmetry breaking occurs. 
This corresponds to the RS phase of the ladder model. 
When $\Jperp<|J+2K_{\rm ch}|$, the pseudospin model \eqref{eq:H_XY} exhibits a variety of $Z_2$ symmetry breaking: 
an antiferromagnetic order along the $x$ axis ($J>2|K_{\rm ch}|$) or the $y$ axis ($2K_{\rm ch}>|J|$) 
and a ferromagnetic order along the $x$ axis ($J<-2|K_{\rm ch}|$) or the $y$ axis ($2\Kch<-|J|$). 
These correspond to the N\'eel, SVC, SF, and UVC phases [see Figs.\ \ref{fig:ladderstates_chpm}(g,h,i,j)]
because of the relations $\Scal_j^x=(S_{1,j}^z-S_{2,j}^z)/2$ and $\Scal_j^y=(\Sv_{1,j}\times\Sv_{2,j})^z$. 
The RS phase for $\Jperp>|J+2K_{\rm ch}|$ exhibits dominant N\'eel, SVC, SF, or UVC correlation 
depending on which of $J$ and $2\Kch$ is dominant and which sign the dominant one has. 
This analysis leads to the phase diagram in Fig.\ \ref{fig:phases_chpm_Jpm_Ising}. 
Focusing on the case of $J>0$, this phase diagram can be converted into the one in the $\Jperp$-$\Kch$ plane in Fig.\ \ref{fig:phases_chpm_Ising}, 
where some similarities with Figs.\ \ref{fig:phases_chpm} and \ref{fig:pd_iDMRG_Exact} are found in view of dominant correlation in the RS phase. 
We note that a crossover between dominant SVC and SF correlations as in Fig.\ \ref{fig:phases_chpm_Jpm_Ising} has also been found 
in the $J$-$\Jperp$-$K$ ladder [Fig.\ \ref{fig:laddermodel_chpm}(b)] with $J=\Jperp<0$ and $K>0$ \cite{Laeuchli03,Totsuka12}.

%%%%%%%%%%%%%%%%%%%%%%%%%%%%%%%%%%%%%%%%%%%%%%%%
\section{Bosonization of Four-Spin Interactions}\label{sec:fourspin_bos}
%%%%%%%%%%%%%%%%%%%%%%%%%%%%%%%%%%%%%%%%%%%%%%%%

\newcommand{\Rrm}{{\rm R}}
\newcommand{\Lrm}{{\rm L}}

Here we provide field-theoretical expressions of the four-spin interactions $H_{\rm LL}$ and $H_{\rm DD}$ in Eq.\ \eqref{eq:CCI_DDI}. 
It is convenient to express these interactions first by the non-Abelian bosonization \cite{Gogolin04}, 
which manifestly respects the SU$(2)$ symmetry of the system. 
Here we closely follow the convention of Ref.\ \cite{Metavitsiadis17}, where similar calculations have been done (see also Ref.\ \cite{Capponi13}). 
In the non-Abelian bosonization, the spin operator is decomposed into uniform and staggered components as 
\begin{equation}
 \Sv_{n,j} = a\qty[ \bm{J}_n(x)+(-1)^j \Omega \bm{n}_n(x) ], 
\end{equation}
where $\Omega$ corresponds to $\ab/a$ in our convention. 
The uniform component is given by the sum of the chiral SU$(2)$ currents as $\bm{J}_n(x)=\bm{J}_{n,\Rrm}(x)+\bm{J}_{n,\Lrm}(x)$. 
By applying operator product expansions, the four-spin interactions are expressed as \cite{Metavitsiadis17}
\begin{subequations}
\begin{align}
 H_{\rm LL} 
 &=\int dx [36\gamma^2 a\Omega^2 O_\epsilon(x) \notag\\
 &~~+3\gamma^2 a^2 \Omega^2 (1+a\Omega^2) O_a(x)+\dots ], \label{eq:HLL_bos}\\
 H_{\rm DD}
 &=\int dx [ 12\gamma^2 a\Omega^2 O_\epsilon \notag\\
 &~~+ 2\gamma^2 a(1+a\Omega^2)^2  \qty( O_b(x)-O(x) )+\dots ],\label{eq:HDD_bos}
\end{align}
\end{subequations}
where $\gamma=1/(2\pi)$, 
\begin{subequations}
\begin{align}
 O_\epsilon(x)&=\epsilon_1(x)\epsilon_2(x),\\
 O_a(x)&=\Jv_{1,\Lrm}(x)\cdot\Jv_{1,\Rrm}(x)+\Jv_{2,\Lrm}(x)\cdot\Jv_{2,\Rrm}(x),\\
 O_b(x)&=\Jv_{1,\Lrm}(x)\cdot\Jv_{2,\Rrm}(x)+\Jv_{1,\Rrm}(x)\cdot\Jv_{2,\Lrm}(x),\\
 O(x)&=\Jv_{1,\Lrm}(x)\cdot\Jv_{2,\Lrm}(x)+\Jv_{1,\Rrm}(x)\cdot\Jv_{2,\Rrm}(x),
\end{align}
\end{subequations}
and the dimer operator $\epsilon_n(x)~(n=1,2)$ corresponds to $\sin(2\sqrt{\pi}\phi_n)$ in our convention. 
Henceforth, we neglect $O(x)$, which is decoupled from the other perturbations in the renormalization group flow. 
We find that both $H_{\rm LL}$ and $H_{\rm DD}$ have the strongly relevant perturbation $O_\epsilon(x)$, 
and in total  give the contributions $\mp 24(2\gamma a\Omega)^2\Kch/(2a)$ to $g_\pm$ in Eq.\ \eqref{eq:coeff_cos_gpm}. 
Differences between $H_{\rm LL}$ and $H_{\rm DD}$ can be found in the marginal perturbations. 
In particular, the inter-chain current-current interaction $O_b(x)$ appears only in $H_{\rm DD}$, 
and this interaction is expected to be responsible for the difference between 
the ladder models of the present study and 
Refs.\ \cite{Nersesyan97, Hijii09, Hijii13, Takayoshi10, Ogino21nvbs,Ogino21spt, ChenJH17,ChenJH19,Robinson19, RenKonik24}, 
as discussed at the end of Sec.\ \ref{sec:bos_phases}. 
In the Abelian bosonization, this interaction can be expressed as \cite{Furukawa12}
\begin{equation}
\begin{split}
 O_b &= - \frac{b_1^2}{2a^2} \cos(2\sqrt{2\pi}\phi_+) \cos(\sqrt{2\pi}\theta_-)\\
 &~~+\frac{1}{8\pi} \sum_{\nu=\pm} \nu \qty[ 2 \qty(\partial_x\phi_\nu)^2 -\frac12 \qty(\partial_x\theta_\nu)^2].
\end{split}
\end{equation}
We therefore find the contribution of the CCI to $\gamma_1$ in Eq.\ \eqref{eq:coeff_gamma1} with $C=8\gamma^2 (1+a\Omega^2)^2$. 

%%%%%%%%%%%%%%%%%%%%%%%%%%%%%%%%%%%%%%%%%%%%%
%\section*{References}
%%%%%%%%%%%%%%%%%%%%%%%%%%%%%%%%%%%%%%%%%%%%%

\bibliography{references}
%\begin{thebibliography}{99}
%\end{thebibliography}

\end{document}